\documentclass[a4paper]{article}
\usepackage{authblk}
\usepackage[margin=2.5cm]{geometry}
\usepackage{mathtools}
\usepackage{amssymb}
\usepackage{amsmath}
\usepackage{xcolor}
\usepackage{biblatex}
\usepackage{booktabs}
\usepackage{float}
\usepackage{tabulary}
\usepackage{tabularx}
\usepackage{array}
\usepackage[title]{appendix}

\title{Shirokov realizations of low dimensional Lie algebras}
\author[1]{Severin Pošta}
\affil[1]{Department of Mathematics, Faculty of Nuclear Sciences and Physical Engineering, Czech Technical University in Prague, Trojanova 13, 120~00 Prague, Czech Republic}
\date{\today}

\begin{document}

\maketitle

\begin{abstract}
We compute the transitive realizations for the low dimensional cases of real Lie algebras up to dimension four using Shirokov's method. First, the generic realizations are given, then, making use of the known list of subalgebras, nongeneric realizations are computed. The result is compared with the known classification of Popovych et al.
\end{abstract}

\section{Introduction}
As it is  well known, it was Sophus Lie himself who was the first who began with the systematic construction of realizations (i. e. representations by vector fields) of Lie algebras.
Nevertheless, the problem of construction of all possible realizations is still unsolved for many important cases of Lie algebras. The classification and description of all realizations of low dimensional Lie algebras is of great interest \cite{Popovych} and has many possible applications. For example, one can use realizations for the integration of ordinary differential
equations, to perform group classification of partial differential
equations, to find differential invariants and construct physical models with various symmetries. It is also a fundamental mathematical problem that is interesting and significant on its own.
The classical methods of construction of realizations of given Lie algebra usually require solving of nontrivial systems of partial differential equations. For a brief overview of other construction methods, you may see, e.g. \cite{Nesterenko_2025}.
One of the very promising methods of construction of transitive realizations that is algebraic and does not require solving differential equations was proposed by Shirokov \cite{Shirokov}.
The main idea of this method is to construct vector fields as duals to differential one-forms, which can be obtained using adjoint representations. The usefulness and practical applicability of this method has already been demonstrated in physically interesting cases \cite{Nesterenko_2016}.

In this article, we examine what happens if we apply Shirokov's method on the low dimensional real Lie algebras up to dimension four. 
The method of Shirokov constructs invariant vector fields from the structure constants by matrix exponentiation and matrix inversion in second canonical coordinates. Once the invariant fields on the local Lie group are known, further transitive realizations can be obtained by projecting to spaces of right cosets associated with subalgebras. The subalgebra representatives used here are taken from Patera and Winternitz, who classified subalgebras of real Lie algebras of dimension at most four \cite{PateraWinternitz}.

For real Lie algebras up to dimension four, Popovych et al. constructed a complete list of inequivalent realizations in vector fields on spaces with an arbitrary finite number of variables \cite{Popovych}. 
The classification of faithful realizations was done up to weak equivalence. The aim of the present paper is different. The paper compares obtained homogeneous realizations with the entries in the Popovych's tables. Unlike the classification of Popovych, which lists faithful realizations, we also record the unfaithful homogeneous realizations produced by Shirokov's projection method; these rows can be interpreted as effective realizations of quotient algebras and indicate precisely which isotropy subalgebras contain nonzero ideals.

Although unfaithful realizations naturally do not have the same significance as faithful ones, there are good reasons to keep them. Patera and Winternitz motivate subalgebra classifications partly by physical uses: subgroup/subalgebra structure is relevant to symmetry breaking, induced representations, choices of representation bases, quantum numbers, invariant operators, and contractions. In such contexts, one often cares about the subgroup or isotropy itself, not only about the effective quotient action. A symmetry algebra can contain an ideal acting trivially on a particular reduced space, and this can be still meaningful geometric information.

The subalgebras are taken modulo inner automorphisms, as in Patera--Winternitz \cite{PateraWinternitz}. 
This is the preferred, natural equivalence for the Shirokov homogeneous construction, since inner-conjugate
subalgebras define locally equivalent homogeneous spaces and hence equivalent projected
Shirokov realizations. In the comparison with Popovych, one must further factor by
weak equivalence, which also allows arbitrary automorphisms of the abstract Lie algebra. In other words,
Shirokov’s equivalence is stricter than Popovych’s weak equivalence. Consequently, different Shirokov rows in our tables may correspond to the same Popovych row.

The structure of the paper is as follows. In the section \ref{shirokovsmethod} we summarize the the practical part of Shirokov's algorithm for constructing transitive realizations. The main part of the paper is in the section \ref{listofalgebrasandtheirrealizations}, where we construct Shirokov realizations for all indecomposable Lie algebras up to dimension 4. Cases of decomposable Lie algebras are briefly discussed in the appendix \ref{decomposable}.

\section{Shirokov's method}
\label{shirokovsmethod}
Let us consider a real Lie algebra $L$, i. e., $L$ is an $n$-dimensional real vector space with a bilinear antisymmetric operation
$[.,.]\colon L\times L\to L$ that satisfies the Jacobi identity. Let us fix the basis $e_1,... ,e_n$ of $L$. The commutation relations of $L$ have the form
\begin{gather*}
  [e_i, e_j]=\sum\limits_{k=1}^n c_{ij}^{k} e_k,\quad i,j=1,...,n,
\end{gather*}
where $c_{ij}^k\in \mathbb{R}$ are the structure constants of $L$. 
Let $M\subset\mathbb{R}^m$, $m\in \mathbb{N}$ be an $m$-dimensional smooth manifold and $\mathop{\mathrm{Vect}}(M)$ denote the Lie algebra of smooth vector fields on $M$. Our aim is to construct a realization of $L$ i. e. a homomorphism of $L$ into $\mathop{\mathrm{Vect}}(M)$. This homomorphism sends each vector $e_i$ to a homogeneous first-order differential operator, namely
\begin{gather*}
e_i \to \xi_i=\xi^{1}_i(x)\partial_1+
\xi^{2}_i(x)\partial_2+...+
\xi^{m}_i(x)\partial_m,
\end{gather*}
where $x=(x_1,...,x_m)$, $\partial_j=\frac{\partial\ }{\partial x_j}$ and the coefficients $\xi^{j}_i$  are smooth functions on $M$.

Shirokov's method for finding the unknowns $\xi^j_i(x)$ is this. First, generic realizations (corresponding
to the primitive action of the local group, i. e. to the case when $m=n$) are constructed. Shirokov starts with the components $\omega^i_j(x)$ of left-invariant differential one-forms dual to the left-invariant vector fields that are given by the formula
\begin{align*}
\omega^i_1(x)&=\delta^i_1,\\ 
\omega^i_j(x)&=\left(\mathop{\mathrm{exp}}(-x_1\mathop{\mathrm{ad}}\nolimits_{e_1}) ... \mathop{\mathrm{exp}}(-x_{j-1}\mathop{\mathrm{ad}}\nolimits_{e_{j-1}}) \right)^i_j, \quad  1\leq i \leq n,\; 1<j\leq n.
\end{align*}
The $j$-th component $\xi^j_i(x)$ of the left-invariant vector field $\xi_i$ is then found as the $(i,j)$-th element of the inverse of the square matrix $(\omega^k_l(x))_{k,l=1}^n$, i. e. 
\begin{equation*} 
\xi^j_i(x)=\left((\omega^k_l(x))^{-1}\right)^j_i,\quad  i,j=1,...,n.
\end{equation*}
Other realizations that correspond to the transitive group action (the case when $m<n$) can be obtained as projections of the generic realizations to the spaces of variables complementary to  subalgebras of $L$. For such construction the knowledge of subalgebras of $L$ is needed.
For a subalgebra $K \subset L$ of dimension $r=n-m$ we choose a basis of $L$ such that \begin{equation*} L=\langle e_1,...,e_m\rangle \oplus K,
\end{equation*} where $\langle ..\rangle$ denotes the real linear span, and compute the left-invariant fields in coordinates adapted to this basis. The realization is obtained by restricting these fields to functions of the complementary coordinates $x_1,...,x_m$. This effectively means formal substitution $\partial_k \to 0$ for $k=m+1,...,n$.

\section{List of algebras and their Shirokov realizations}
\label{listofalgebrasandtheirrealizations}

In this section we give a list of low dimensional real Lie algebras together with their Shirokov realizations obtained by the method described above. We take the list of algebras from Šnobl's book \cite{Snobl} and use their Lie algebra notation. Since we are comparing the obtained Shirokov realizations with Popovych \cite{Popovych}, we use almost always the Popovych commutation relations. This is the natural choice, since the realizations are compared to Popovych ones, and, moreover, Popovych's notation, basis choices and commutation relations coincide with Mubarakzyanov’s ones. The only exception to this rule among the indecomposable algebras is the exceptional diagonal case $s_{4,3}(a,a)$, $a\neq0,1$ (table \ref{s43scalar}). There we keep the natural basis and do not normalize to Popovych's representative.

The results are summarized into tables. If necessary, various notes are provided. Each table contains the name of the algebra, its dimension, the commutation relations we work in, the commutation relations used in \cite{PateraWinternitz}, \cite{Popovych}, \cite{Mubarakzyanov} (if they differ from ours) and the overview of its Shirokov realizations. For each Shirokov realization, the corresponding Popovych realization from \cite{Popovych} is given. Sometimes the Popovych realization agrees exactly with our formulas. Sometimes, we need to use additional automorphisms of the algebra, denoted usually by $\tilde e_i$, and/or local coordinate changes, denoted by $y_i$, to put a Shirokov realization into Popovych's “normal form”.

The coordinates are local to each corresponding realization. Thus, in each row, $\partial_i$ means differentiation with respect to the $i$-th coordinate on the quotient space associated with the chosen complement.

The list of algebras and their realizations follow. The first two subsections are commented in more detail, the others briefly.

\subsection{The algebra $n_{1,1}$}

This is the unique one-dimensional Lie algebra, denoted $n_{1,1}$ in Šnobl's classification \cite{Snobl}. It is denoted as $A_1$ by both Patera-Winternitz \cite{PateraWinternitz}, and Popovych \cite{Popovych}. It contains no parameters.

The only proper subalgebra is the trivial one, $\{0\}$, which corresponds to the generic Shirokov realization $e_{1}=\partial_{1}$. Thus, there are no non-trivial nongeneric realizations to construct.
The generic realization exactly matches the single realization $R(A_1, 1)$ presented by Popovych \cite{Popovych}. The results are summarized in table \ref{n11}.

\begin{table}[H]
\centering
\small
\begin{tabularx}{\linewidth}{llX}
\toprule
Algebra & dim. & Nonzero commutation relations  \\
$n_{1,1}$ & $1$ & none  \\
\bottomrule
\end{tabularx}
\begin{tabularx}{\linewidth}{llX}
Patera-Winternitz & Popovych & Mubarakzyanov \\
$A_1$ & $A_1$  & $g_1$ \\
\bottomrule
\end{tabularx}
\medskip
\begin{tabularx}{\linewidth}{llclcXr}
\toprule
Subalgebra & Complement & dim. & Kernel & Faithful? & Shirokov realization & Popovych realization \\
\midrule
$\{0\}$ & $\langle e_1\rangle$ & $1$ & $\{0\}$ & yes & $e_1=\partial_1$ & $R(A_1,1)$ \\
\bottomrule
\end{tabularx}
\caption{The algebra $n_{1,1}$}
\label{n11}
\end{table}

\subsection{The algebra $s_{2,1}$}

The algebra $\mathrm{aff}(1)$, denoted $s_{2,1}$ in Šnobl's classification \cite{Snobl}, contains no parameters. In Patera-Winternitz \cite{PateraWinternitz} this algebra is denoted as $A_2$, in Popovych \cite{Popovych} it is denoted $A_{2.1}$. In Mubarakzyanov \cite{Mubarakzyanov}, this algebra is denoted $g_{2}$.
We use Popovych/Mubarakzyanov basis and the commutation relation \begin{equation*}
    [e_1,e_2]=e_1.
\end{equation*}
Patera-Winternitz use a different basis, which we denote by $\{p_1,p_2\}$, namely they have
\begin{equation*}
    [p_1,p_2]=p_2.
\end{equation*}
Šnobl uses the basis that we denote $\{f_1,f_2\}$. In this basis he has
\begin{equation*}
    [f_2,f_1]=f_1.
\end{equation*}
Thus, we have the following conversion of these basis vectors:
\begin{equation*}
    p_1=f_2,\quad p_2=f_1,\quad e_1=f_1=p_2,\quad e_2=-f_2=-p_1.
\end{equation*}
In our basis $\left\{e_1,e_2\right\}$, the Shirokov construction gives a faithful generic realization
\begin{equation*}
    e_1=\partial_1,\quad  e_2=x_1 \partial_1 +\partial_2.
\end{equation*}
This realization exactly agrees with the Popovych realization $R(A_{2.1},1)$. 

The algebra has two nontrivial subalgebras, nonequivalent up to inner automorphism: $\langle e_1 \rangle$ and $\langle e_2\rangle$. 
The basis element $e_1$ spans the derived algebra, which is an ideal. If we quotient by this ideal (projecting the generic realization to the space of functions of the complementary variable $x_2$ by dropping the $\partial_{1}$ terms), we obtain the unfaithful, nongeneric realization 
\begin{equation*}
e_1 = 0,\quad  e_2 = \partial_{2}.
\end{equation*}

Regarding the subalgebra $\langle e_2\rangle$, the complementary space is parameterized by the coordinate $x_1$. We project the generic realization to the space of functions of $x_1$ by dropping the $\partial_{2}$ terms. This yields the following faithful nongeneric realization:
\begin{equation*}
e_{1}=\partial_{1},\quad e_{2}=x_{1}\partial_{1}. 
\end{equation*}
This agrees with the Popovych's $R(A_{2.1},2)$.
The results are summarized in the table \ref{s21}.

\begin{table}[H]
\centering
\small
\begin{tabularx}{\linewidth}{lllX}
\toprule
Algebra & Also & dim. & Nonzero commutation relations  \\
$s_{2,1}$ & $\mathrm{aff}(1)$ & $2$ & $[e_1,e_2]=e_1$  \\
\bottomrule
\end{tabularx}
\begin{tabularx}{\linewidth}{lllX}
Patera-Winternitz & Šnobl & Popovych  & Mubarakzyanov \\
$A_2$ &  $f_1=e_1$, $f_2=-e_2$   & $A_{2.1}$ & $g_{2}$ \\
$p_1=-e_2$, $p_2=e_1$   & $[f_2,f_1]=f_1$ &  &  \\
$[p_1,p_2]=p_2$ &  &  \\
\bottomrule
\end{tabularx}
\medskip
\begin{tabularx}{\linewidth}{llclcXr}
\toprule
Subalgebra & Complement & dim. & Kernel & Faithful? & Shirokov realization & Popovych realization \\
\midrule
$\{0\}$ & $\langle e_1,e_2\rangle$ & $2$ & $\{0\}$ & yes & $e_1=\partial_1$, $e_2=x_1\partial_1+\partial_2$ & $R(A_{2.1},1)$ \\
$\langle e_1\rangle$ & $\langle e_2 \rangle$ & $1$ & $\langle e_1\rangle$ & no & $e_1=0$, $e_2=\partial_2$ &  \\
$\langle e_2\rangle$ & $\langle e_1\rangle$ & $1$ & $\{0\}$ & yes & $e_1=\partial_1$, $e_2=x_1\partial_1$ & $R(A_{2.1},2)$ \\
\bottomrule
\end{tabularx}
\caption{The algebra $s_{2,1}$}
\label{s21}
\end{table}

\subsection{The algebra $n_{3,1}$}

The realizations of the Heisenberg (Heisenberg-Weyl) algebra $n_{3,1}$, also sometimes denoted by $h_3$ or $h_1$, are summarized in table \ref{n31}. Here \begin{equation*}
    v_\phi=e_2\cos\phi+e_3\sin\phi,\qquad
    w_\phi=-e_2\sin\phi+e_3\cos\phi,
    \qquad 0\leq \phi<\pi .
\end{equation*}
The generic realization (corresponding to the $\{0\}$-subalgebra) agrees exactly with Popovych's $R(A_{3.1},1)$.
For the faithful two-dimensional realization corresponding to $\langle v_\phi\rangle$, one convenient equivalence with Popovych's $R(A_{3.1},3)$ is obtained by using the automorphic basis
\begin{equation*}
    \tilde e_1=-e_1,\qquad \tilde e_2=w_\phi,
    \qquad \tilde e_3=v_\phi,
\end{equation*}
and the coordinate change $y_1=-x_1$, $y_2=x_2$.

\begin{table}[H]
\centering
\small
\begin{tabularx}{\linewidth}{lllX}
\toprule
Algebra & Also & dim. & Nonzero commutation relations  \\
$n_{3,1}$ & $h_3$, Bianchi type II & $3$ & $[e_2,e_3]=e_1$  \\
\bottomrule
\end{tabularx}
\begin{tabularx}{\linewidth}{llX}
Patera-Winternitz & Popovych  & Mubarakzyanov \\
$A_{3,1}$ &  $A_{3.1}$ & $g_{3,1}$ \\
\bottomrule
\end{tabularx}
\begin{tabulary}{\linewidth}{LLCLCLR}
\toprule
Subalgebra & Complement & dim. & Kernel & Faithful? & Shirokov realization & Popovych realization \\
\midrule
$\{0\}$
& $\langle e_1,e_2,e_3\rangle$
& $3$ & $\{0\}$ & yes
& $e_1=\partial_1$, $e_2=\partial_2$, $e_3=x_2\partial_1+\partial_3$
& $R(A_{3.1},1)$ \\
\addlinespace
$\langle e_1\rangle$
& $\langle e_2,e_3\rangle$
& $2$ & $\langle e_1\rangle$ & no
& $e_1=0$, $e_2=\partial_1$, $e_3=\partial_2$
&  \\
\addlinespace
$\langle v_\phi\rangle$
& $\langle e_1,w_\phi\rangle$
& $2$ & $\{0\}$ & yes
& $e_1=\partial_1$,
 $e_2=-x_2\cos\phi\,\partial_1-\sin\phi\,\partial_2$,
 $e_3=-x_2\sin\phi\,\partial_1+\cos\phi\,\partial_2$
& $R(A_{3.1},3)$ \\
\addlinespace
$\langle e_1,v_\phi\rangle$
& $\langle w_\phi\rangle$
& $1$ & $\langle e_1,v_\phi\rangle$ & no
& $e_1=0$, $e_2=-\sin\phi\,\partial_1$, $e_3=\cos\phi\,\partial_1$
&  \\
\bottomrule
\end{tabulary}
\caption{The algebra $n_{3,1}$}
\label{n31}
\end{table}

\subsection{The algebra $s_{3,1}(a)$, $a\neq 1$}

Here $0<|a|\leq 1$, $a\neq 1$. This is the Lie algebra of anisotropic dilations and translations. When $a=-1$, this is the Lie algebra of pseudo-Euclidean group $E(1,1)$, also known as Poincaré algebra in 1+1 dimensions $p(1,1)$. The realizations are summarized in the table \ref{s31generic}. The parameter $\varepsilon=\pm1$.

For the row $\langle e_1+\varepsilon e_2\rangle$, the equivalence to Popovych's $R(A^a_{3.4},3)$ is given by
\begin{equation*}
    y_1=x_1,
    \qquad
    y_2=-\varepsilon e^{(1-a)x_2}.
\end{equation*}

\begin{table}[H]
\centering
\small
\begin{tabulary}{\linewidth}{LLLL}
\toprule
Algebra & Also & dim. & Nonzero commutation relations  \\
$s_{3,1}(a)$ & Bianchi type VI; Bianchi type VI$_0$ or $E(1,1)$ or $p(1,1)$ when $a=-1$ & $3$ &   $[e_1,e_3]=e_1$, $[e_2,e_3]=a e_2$ \\
\bottomrule
\end{tabulary}
\begin{tabularx}{\linewidth}{llll}
Patera-Winternitz & Popovych  & Šnobl  & Mubarakzyanov \\
$A_{3,4}$ if $a=-1$, $A^a_{3,5}$ if $0<|a|<1$ & $A_{3.4}^a$ & $f_1=e_1$, $f_2=e_2$, $f_3=-e_3$ & $g_{3,4}^h$, $h=a$ \\
& & $[f_3,f_1]=f_1$, $[f_3,f_2]=a f_2$ & \\
\bottomrule
\end{tabularx}
\medskip
\begin{tabulary}{\linewidth}{LLCLCLR}
\toprule
Subalgebra & Complement & dim. & Kernel & Faithful? & Shirokov realization & Popovych realization \\
\midrule
$\{0\}$
& $\langle e_1,e_2,e_3\rangle$
& $3$ & $\{0\}$ & yes
& $e_1=\partial_1$, $e_2=\partial_2$, $e_3=x_1\partial_1+a x_2\partial_2+\partial_3$ 
& $R(A^a_{3.4},1)$ \\
\addlinespace
$\langle e_1\rangle$
& $\langle e_2,e_3\rangle$
& $2$ & $\langle e_1\rangle$ & no
& $e_1=0$, $e_2=\partial_1$, $e_3=a x_1\partial_1+\partial_2$
&  \\
\addlinespace
$\langle e_2\rangle$
& $\langle e_1,e_3\rangle$
& $2$ & $\langle e_2\rangle$ & no
& $e_1=\partial_1$, $e_2=0$, $e_3=x_1\partial_1+\partial_2$
&  \\
\addlinespace
$\langle e_3\rangle$
& $\langle e_1,e_2\rangle$
& $2$ & $\{0\}$ & yes
& $e_1=\partial_1$, $e_2=\partial_2$, $e_3=x_1\partial_1+a x_2\partial_2$
& $R(A^a_{3.4},2)$ \\
\addlinespace
$\langle e_1+\varepsilon e_2\rangle$
& $\langle e_1,e_3\rangle$
& $2$ & $\{0\}$ & yes
& $e_1=\partial_1$, $e_2=-\varepsilon e^{(1-a)x_2}\partial_1$, $e_3=x_1\partial_1+\partial_2$
& $R(A^a_{3.4},3)$ \\
\addlinespace
$\langle e_1,e_2\rangle$
& $\langle e_3\rangle$
& $1$ & $\langle e_1,e_2\rangle$ & no
& $e_1=0$, $e_2=0$, $e_3=\partial_1$
&  \\
\addlinespace
$\langle e_1,e_3\rangle$
& $\langle e_2\rangle$
& $1$ & $\langle e_1\rangle$ & no
& $e_1=0$, $e_2=\partial_1$, $e_3=a x_1\partial_1$
&  \\
\addlinespace
$\langle e_2,e_3\rangle$
& $\langle e_1\rangle$
& $1$ & $\langle e_2\rangle$ & no
& $e_1=\partial_1$, $e_2=0$, $e_3=x_1\partial_1$
&  \\
\bottomrule
\end{tabulary}
\caption{The algebra $s_{3,1}(a)$, $a\neq 1$}
\label{s31generic}
\end{table}

\subsection{The special case $s_{3,1}(1)$}

For $a=1$, the algebra $s_{3,1}(a)$ becomes $s_{3,1}(1)$, the Lie algebra of the group of dilations and translations of the 2D plane, and the subalgebra list must be written separately. The results are summarized in the table \ref{s31special}, where 
\begin{equation*}
    v_\phi=e_1\cos\phi+e_2\sin\phi,
    \qquad
    w_\phi=-e_1\sin\phi+e_2\cos\phi,
    \qquad 0\leq \phi<\pi.
\end{equation*}

\begin{table}[H]
\centering
\small
\begin{tabularx}{\linewidth}{llll}
\toprule
Algebra & Also & dim. & Nonzero commutation relations  \\
$s_{3,1}(1)$ & $DT_2$ & $3$ &   $[e_1,e_3]=e_1$, $[e_2,e_3]=e_2$ \\
\bottomrule
\end{tabularx}
\begin{tabularx}{\linewidth}{llll}
Patera-Winternitz & Popovych  & Šnobl  & Mubarakzyanov \\
$A_{3,3}$  & $A_{3.3}$ & $f_1=e_1$, $f_2=e_2$, $f_3=-e_3$ & $g_{3,3}$ \\
& & $[f_3,f_1]=f_1$, $[f_3,f_2]=f_2$ & \\
\bottomrule
\end{tabularx}
\medskip
\begin{tabulary}{\linewidth}{LLCLCLR}
\toprule
Subalgebra & Complement & dim. & Kernel & Faithful? & Shirokov realization & Popovych realization \\
\midrule
$\{0\}$
& $\langle e_1,e_2,e_3\rangle$
& $3$ & $\{0\}$ & yes
& $e_1=\partial_1$, $e_2=\partial_2$, $e_3=x_1\partial_1+x_2\partial_2+\partial_3$
& $R(A_{3.3},1)$ \\
\addlinespace
$\langle v_\phi\rangle$
& $\langle w_\phi,e_3\rangle$
& $2$ & $\langle v_\phi\rangle$ & no
& $e_1=-\sin\phi\,\partial_1$, $e_2=\cos\phi\,\partial_1$, $e_3=x_1\partial_1+\partial_2$
&  \\
\addlinespace
$\langle e_3\rangle$
& $\langle e_1,e_2\rangle$
& $2$ & $\{0\}$ & yes
& $e_1=\partial_1$, $e_2=\partial_2$, $e_3=x_1\partial_1+x_2\partial_2$
& $R(A_{3.3},2)$ \\
\multicolumn{7}{c}{(continued on next page)}\end{tabulary}\end{table}\begin{table}[H]\centering\small\begin{tabulary}{\linewidth}{LLCLCLR} Subalgebra & Complement & dim. & Kernel & Faithful? & Shirokov realization & Popovych realization \\ \midrule \multicolumn{7}{c}{(continued from previous page)} \\
$\langle e_1,e_2\rangle$
& $\langle e_3\rangle$
& $1$ & $\langle e_1,e_2\rangle$ & no
& $e_1=0$, $e_2=0$, $e_3=\partial_1$
&  \\
\addlinespace
$\langle e_3,v_\phi\rangle$
& $\langle w_\phi\rangle$
& $1$ & $\langle v_\phi\rangle$ & no
& $e_1=-\sin\phi\,\partial_1$, $e_2=\cos\phi\,\partial_1$, $e_3=x_1\partial_1$
&  \\
\bottomrule
\end{tabulary}
\caption{The algebra $s_{3,1}(1)$}
\label{s31special}
\end{table}

\subsection{The algebra $s_{3,2}$}

The algebra $s_{3,2}$ is the three-dimensional solvable Lie algebra with two-dimensional Abelian nilradical and a nontrivial Jordan block. In the Bianchi notation it corresponds to Bianchi type IV. 
The faithful generic realization agrees with Popovych's $R(A_{3.2},1)$. For the row corresponding to the subalgebra $\langle e_2\rangle$, the coordinate change
\begin{equation*}
    y_1=x_1,\qquad y_2=-x_2
\end{equation*}
brings the realization to Popovych's 
form $R(A_{3.2},3)$. The results are summarized in table \ref{s32}.

\begin{table}[H]
\centering
\small
\begin{tabularx}{\linewidth}{llll}
\toprule
Algebra & Also & dim. & Nonzero commutation relations \\
$s_{3,2}$ & Bianchi type IV & $3$ & $[e_1,e_3]=e_1$, $[e_2,e_3]=e_1+e_2$ \\
\bottomrule
\end{tabularx}
\begin{tabularx}{\linewidth}{llll}
Patera-Winternitz & Popovych & Šnobl & Mubarakzyanov \\
$A_{3,2}$ & $A_{3.2}$ & $f_1=e_1$, $f_2=e_2$, $f_3=-e_3$ & $g_{3,2}$ \\
 &  & $[f_3,f_1]=f_1$, $[f_3,f_2]=f_1+f_2$ &  \\
\bottomrule
\end{tabularx}
\medskip
\begin{tabulary}{\linewidth}{LLCLCLR}
\toprule
Subalgebra & Complement & dim. & Kernel & Faithful? & Shirokov realization & Popovych realization \\
\midrule
$\{0\}$ & $\langle e_1,e_2,e_3\rangle$ & $3$ & $\{0\}$ & yes & $e_1=\partial_1$, $e_2=\partial_2$, $e_3=(x_1+x_2)\partial_1+x_2\partial_2+\partial_3$ & $R(A_{3.2},1)$ \\
\addlinespace
$\langle e_1\rangle$ & $\langle e_2,e_3\rangle$ & $2$ & $\langle e_1\rangle$ & no & $e_1=0$, $e_2=\partial_1$, $e_3=x_1\partial_1+\partial_2$ &  \\
\addlinespace
$\langle e_2\rangle$ & $\langle e_1,e_3\rangle$ & $2$ & $\{0\}$ & yes & $e_1=\partial_1$, $e_2=-x_2\partial_1$, $e_3=x_1\partial_1+\partial_2$ & $R(A_{3.2},3)$ \\
\addlinespace
$\langle e_3\rangle$ & $\langle e_1,e_2\rangle$ & $2$ & $\{0\}$ & yes & $e_1=\partial_1$, $e_2=\partial_2$, $e_3=(x_1+x_2)\partial_1+x_2\partial_2$ & $R(A_{3.2},2)$ \\
\addlinespace
$\langle e_1,e_2\rangle$ & $\langle e_3\rangle$ & $1$ & $\langle e_1,e_2\rangle$ & no & $e_1=0$, $e_2=0$, $e_3=\partial_1$ &  \\
\addlinespace
$\langle e_1,e_3\rangle$ & $\langle e_2\rangle$ & $1$ & $\langle e_1\rangle$ & no & $e_1=0$, $e_2=\partial_1$, $e_3=x_1\partial_1$ &  \\
\bottomrule
\end{tabulary}
\caption{The algebra $s_{3,2}$}
\label{s32}
\end{table}

\subsection{The algebra $s_{3,3}(a)$}

Here $a \geq 0$. This is the solvable algebra with two-dimensional Abelian nilradical on which the nonnilpotent generator acts by a real Jordan form corresponding to a complex pair of eigenvalues; in the Bianchi notation it is type VII$_a$. The special case $a=0$ is the Euclidean algebra $e(2)$, the Lie algebra of the Euclidean group
$E(2)$.
For the row $\langle e_2\rangle$ we use a conjugate representative of the one-dimensional subalgebra in the Abelian ideal and the local coordinate obtained from the raw adapted coordinate by
\begin{equation*}
    y_1=x_1,\qquad y_2=-\tan x_2.
\end{equation*}
With this convention the row is comparable directly with Popovych's $R(A^a_{3.5},3)$. The results are summarized in table \ref{s33}.

\begin{table}[H]
\centering
\small
\begin{tabulary}{\linewidth}{LLLL}
\toprule
Algebra & Also & dim. & Nonzero commutation relations \\
$s_{3,3}(a)$ & Bianchi type VII$_a$; Bianchi type VII$_0$ or $e(2)$ for $a=0$ & $3$ & $[e_1,e_3]=a e_1-e_2$, $[e_2,e_3]=e_1+a e_2$ \\
\bottomrule
\end{tabulary}
\begin{tabulary}{\linewidth}{LLLL}
Patera-Winternitz & Popovych & Šnobl & Mubarakzyanov \\
$A_{3,7}^{a}$ for $a>0$, $A_{3,6}$ for $a=0$ & $A_{3.5}^{a}$ & $f_1=e_1$, $f_2=e_2$, $f_3=-e_3$ & $g_{3,5}^{a}$ \\
 &  & $[f_3,f_1]=a f_1-f_2$, $[f_3,f_2]=f_1+a f_2$ & \\
\bottomrule
\end{tabulary}
\medskip
\begin{tabulary}{\linewidth}{LLCLCLR}
\toprule
Subalgebra & Complement & dim. & Kernel & Faithful? & Shirokov realization & Popovych realization \\
\midrule
$\{0\}$ & $\langle e_1,e_2,e_3\rangle$ & $3$ & $\{0\}$ & yes & $e_1=\partial_1$, $e_2=\partial_2$, $e_3=(a x_1+x_2)\partial_1+(-x_1+a x_2)\partial_2+\partial_3$ & $R(A^a_{3.5},1)$ \\
\addlinespace
$\langle e_2\rangle$ & $\langle e_1,e_3\rangle$ & $2$ & $\{0\}$ & yes & $e_1=\partial_1$, $e_2=x_2\partial_1$, $e_3=(a-x_2)x_1\partial_1-(1+x_2^2)\partial_2$ & $R(A^a_{3.5},3)$ \\
\addlinespace
$\langle e_3\rangle$ & $\langle e_1,e_2\rangle$ & $2$ & $\{0\}$ & yes & $e_1=\partial_1$, $e_2=\partial_2$, $e_3=(a x_1+x_2)\partial_1+(-x_1+a x_2)\partial_2$ & $R(A^a_{3.5},2)$ \\
\addlinespace
$\langle e_1,e_2\rangle$ & $\langle e_3\rangle$ & $1$ & $\langle e_1,e_2\rangle$ & no & $e_1=0$, $e_2=0$, $e_3=\partial_1$ &  \\
\bottomrule
\end{tabulary}
\caption{The algebra $s_{3,3}(a)$, $a \geq 0$}
\label{s33}
\end{table}

\subsection{The algebra $\mathrm{sl}(2,\mathbb{R})$}

This is real simple algebra,
also known as Bianchi type VIII.
For comparison with Popovych, the two-dimensional quotient rows use
\begin{equation*}
    y_1=x_1,\qquad y_2=e^{x_2}.
\end{equation*}
The one-dimensional row $\langle e_2,e_3\rangle$ agrees directly with Popovych. The results are summarized in table \ref{sl2r}.

\begin{table}[H]
\centering
\small
\begin{tabularx}{\linewidth}{llll}
\toprule
Algebra & Also & dim. & Nonzero commutation relations \\
$\mathrm{sl}(2,\mathbb{R})$ & Bianchi type VIII & $3$ & $[e_1,e_2]=e_1$, $[e_2,e_3]=e_3$, $[e_1,e_3]=2e_2$ \\
\bottomrule
\end{tabularx}
\begin{tabularx}{\linewidth}{ll}
Patera-Winternitz & Šnobl \\
$A_{3,8}$ & $f_1=e_1$, $f_2=2e_2$, $f_3=-e_3$ \\
$p_1=e_1$, $p_2=e_2$, $p_3=-e_3$ & $[f_1,f_2]=2f_1$, $[f_1,f_3]=-f_2$, $[f_2,f_3]=2f_3$ \\
$[p_1,p_2]=p_1$, $[p_1,p_3]=-2p_2$, $[p_2,p_3]=p_3$ & \\
\bottomrule
\end{tabularx}
\medskip
\begin{tabulary}{\linewidth}{LLCLCLR}
\toprule
Subalgebra & Complement & dim. & Kernel & Faithful? & Shirokov realization & Popovych realization \\
\midrule
$\{0\}$ & $\langle e_1,e_2,e_3\rangle$ & $3$ & $\{0\}$ & yes & $e_1=\partial_1$, $e_2=x_1\partial_1+x_2\partial_2$, $e_3=x_1^2\partial_1+2x_1x_2\partial_2+x_2\partial_3$ & $R(\mathrm{sl}(2,\mathbb{R}),1)$ \\
\addlinespace
$\langle e_1+e_3\rangle$ & $\langle e_1,e_2\rangle$ & $2$ & $\{0\}$ & yes & $e_1=\partial_1$, $e_2=x_1\partial_1+x_2\partial_2$, $e_3=(x_1^2-x_2^2)\partial_1+2x_1x_2\partial_2$ & $R(\mathrm{sl}(2,\mathbb{R}),2)$ \\
\addlinespace
$\langle e_1-e_3\rangle$ & $\langle e_1,e_2\rangle$ & $2$ & $\{0\}$ & yes & $e_1=\partial_1$, $e_2=x_1\partial_1+x_2\partial_2$, $e_3=(x_1^2+x_2^2)\partial_1+2x_1x_2\partial_2$ & $R(\mathrm{sl}(2,\mathbb{R}),3)$ \\
\addlinespace
$\langle e_3\rangle$ & $\langle e_1,e_2\rangle$ & $2$ & $\{0\}$ & yes & $e_1=\partial_1$, $e_2=x_1\partial_1+x_2\partial_2$, $e_3=x_1^2\partial_1+2x_1x_2\partial_2$ & $R(\mathrm{sl}(2,\mathbb{R}),4)$ \\
\addlinespace
$\langle e_2,e_3\rangle$ & $\langle e_1\rangle$ & $1$ & $\{0\}$ & yes & $e_1=\partial_1$, $e_2=x_1\partial_1$, $e_3=x_1^2\partial_1$ & $R(\mathrm{sl}(2,\mathbb{R}),5)$ \\
\bottomrule
\end{tabulary}
\caption{The algebra $\mathrm{sl}(2,\mathbb{R})$}
\label{sl2r}
\end{table}

\subsection{The algebra $\mathrm{so}(3)$}

This is simple Lie algebra, also known as Bianchi type IX.
All one-dimensional subalgebras are conjugate under inner automorphisms. We use $\langle e_2\rangle$ in the table because it gives the usual spherical-coordinate form of the two-dimensional realization in Popovych's list. The results are summarized in table \ref{so3}.

\begin{table}[H]
\centering
\small
\begin{tabularx}{\linewidth}{llll}
\toprule
Algebra & Also & dim. & Nonzero commutation relations \\
$\mathrm{so}(3)$ & Bianchi type IX & $3$ & $[e_2,e_3]=e_1$, $[e_3,e_1]=e_2$, $[e_1,e_2]=e_3$ \\
\bottomrule
\end{tabularx}
\begin{tabularx}{\linewidth}{l}
Patera-Winternitz \\
$A_{3,9}$ \\
\bottomrule
\end{tabularx}
\medskip
\begin{tabulary}{\linewidth}{LLCLCLR}
\toprule
Subalgebra & Complement & dim. & Kernel & Faithful? & Shirokov realization & Popovych realization \\
\midrule
$\{0\}$ & $\langle e_1,e_2,e_3\rangle$ & $3$ & $\{0\}$ & yes & $e_1=-\sin x_1\tan x_2\,\partial_1 -\cos x_1\,\partial_2 +\sin x_1\mathrm{sec}\,x_2\,\partial_3$, $e_2=-\cos x_1\tan x_2\,\partial_1 +\sin x_1\,\partial_2 +\cos x_1\mathrm{sec}\,x_2\,\partial_3$, $e_3=\partial_1$ & $R(\mathrm{so}(3),2)$ \\
\addlinespace
$\langle e_2\rangle$ & $\langle e_1,e_3\rangle$ & $2$ & $\{0\}$ & yes & $e_1=-\sin x_1\tan x_2\,\partial_1 -\cos x_1\,\partial_2$, $e_2=-\cos x_1\tan x_2\,\partial_1 +\sin x_1\,\partial_2$, $e_3=\partial_1$ & $R(\mathrm{so}(3),1)$ \\
\bottomrule
\end{tabulary}
\caption{The algebra $\mathrm{so}(3)$}
\label{so3}
\end{table}

\subsection{The algebra $n_{4,1}$}

The algebra $n_{4,1}$ is the four-dimensional filiform nilpotent Lie algebra.
The parameter $\lambda$ in some subalgebra representatives is the Patera-Winternitz parameter. For the faithful rows in which this parameter occurs, it may be removed from the comparison with Popovych by automorphisms of the algebra, for instance by shifted generators of the form $\tilde e_4=e_4+\lambda e_3$ or $\tilde e_3=e_3+c e_2$, $c$ being suitable constant, when the corresponding adapted complement is changed. The results are summarized in table \ref{n41}.

\begin{table}[H]
\centering
\small
\begin{tabularx}{\linewidth}{lll}
\toprule
Algebra &  dim. & Nonzero commutation relations \\
$n_{4,1}$ &  $4$ & $[e_2,e_4]=e_1$, $[e_3,e_4]=e_2$ \\
\bottomrule
\end{tabularx}
\begin{tabularx}{\linewidth}{lll}
Patera-Winternitz & Popovych & Mubarakzyanov \\
$A_{4,1}$ & $A_{4.1}$ & $g_{4,1}$ \\
\bottomrule
\end{tabularx}
\medskip
\begin{tabulary}{\linewidth}{LLCLCLR}
\toprule
Subalgebra & Complement & dim. & Kernel & Faithful? & Shirokov realization & Popovych realization \\
\midrule
$\{0\}$ & $\langle e_1,e_2,e_3,e_4\rangle$\hphantom{xxx} & $4$ & $\{0\}$ & yes & $e_1=\partial_1$, $e_2=\partial_2$, $e_3=\partial_3$, $e_4=x_2\partial_1+x_3\partial_2+\partial_4$ & $R(A_{4.1},1)$ \\
\addlinespace
$\langle e_1\rangle$ & $\langle e_2,e_3,e_4\rangle$ & $3$ & $\langle e_1\rangle$ & no & $e_1=0$, $e_2=\partial_1$, $e_3=\partial_2$, $e_4=x_2\partial_1+\partial_3$ &  \\
\addlinespace
$\langle e_2\rangle$ & $\langle e_1,e_3,e_4\rangle$ & $3$ & $\{0\}$ & yes & $e_1=\partial_1$, $e_2=-x_3\partial_1$, $e_3=\partial_2$, $e_4=-x_2x_3\partial_1+\partial_3$ & $R(A_{4.1},6)$ \\
\addlinespace
$\langle e_3+\lambda e_1\rangle$ & $\langle e_1,e_2,e_4\rangle$ & $3$ & $\{0\}$ & yes & $e_1=\partial_1$, $e_2=\partial_2$, $e_3=-(\lambda+\tfrac12x_3^2)\partial_1-x_3\partial_2$, $e_4=x_2\partial_1+\partial_3$ & $R(A_{4.1},5)$ \\
\addlinespace
$\langle e_4+\lambda e_3\rangle$ & $\langle e_1,e_2,e_3\rangle$ & $3$ & $\{0\}$ & yes & $e_1=\partial_1$, $e_2=\partial_2$, $e_3=\partial_3$, $e_4=x_2\partial_1+x_3\partial_2-\lambda\partial_3$ & $R(A_{4.1},3)$ \\
\addlinespace
$\langle e_1,e_2\rangle$ & $\langle e_3,e_4\rangle$ & $2$ & $\langle e_1,e_2\rangle$ & no & $e_1=0$, $e_2=0$, $e_3=\partial_1$, $e_4=\partial_2$ &  \\
\addlinespace
$\langle e_1,e_3\rangle$ & $\langle e_2,e_4\rangle$ & $2$ & $\langle e_1\rangle$ & no & $e_1=0$, $e_2=\partial_1$, $e_3=-x_2\partial_1$, $e_4=\partial_2$ &  \\
\addlinespace
$\langle e_2,e_3+\lambda e_1\rangle$ & $\langle e_1,e_4\rangle$ & $2$ & $\{0\}$ & yes & $e_1=\partial_1$, $e_2=-x_2\partial_1$, $e_3=(-\lambda+\tfrac12x_2^2)\partial_1$, $e_4=\partial_2$ & $R(A_{4.1},8)$ \\
\addlinespace
$\langle e_1,e_4+\lambda e_3\rangle$ & $\langle e_2,e_3\rangle$ & $2$ & $\langle e_1\rangle$ & no & $e_1=0$, $e_2=\partial_1$, $e_3=\partial_2$, $e_4=x_2\partial_1-\lambda\partial_2$ &  \\
\addlinespace
$\langle e_1,e_2,e_3\rangle$ & $\langle e_4\rangle$ & $1$ & $\langle e_1,e_2,e_3\rangle$ & no & $e_1=0$, $e_2=0$, $e_3=0$, $e_4=\partial_1$ &  \\
\addlinespace
$\langle e_1,e_2,e_4+\lambda e_3\rangle$ & $\langle e_3\rangle$ & $1$ & $\langle e_1,e_2,e_4+\lambda e_3\rangle$ & no & $e_1=0$, $e_2=0$, $e_3=\partial_1$, $e_4=-\lambda\partial_1$ &  \\
\bottomrule
\end{tabulary}
\caption{The algebra $n_{4,1}$}
\label{n41}
\end{table}

\subsection{The algebra $s_{4,1}$}

This is a solvable four-dimensional algebra with three-dimensional Abelian nilradical. 
In the row $\langle e_3+\lambda e_1\rangle$, Popovych's discrete parameter is $\varepsilon=0$ for $\lambda=0$ and $\varepsilon=1$ for $\lambda\neq0$. In the row $\langle e_1+\lambda e_2,e_3\rangle$, $\lambda\neq0$, the change
\begin{equation*}
    y_1=x_1,\qquad y_2=-\lambda^{-1}e^{x_2}
\end{equation*}
together with an admissible shift of $e_3$ by a suitable multiple of $e_2$ gives Popovych's logarithmic normal form. The results are summarized in table \ref{s41}.

\begin{table}[H]
\centering
\small
\begin{tabularx}{\linewidth}{lll}
\toprule
Algebra &  dim. & Nonzero commutation relations \\
$s_{4,1}$ &   $4$ & $[e_1,e_4]=e_1$, $[e_3,e_4]=e_2$ \\
\bottomrule
\end{tabularx}
\begin{tabularx}{\linewidth}{llll}
Patera-Winternitz & Popovych & Šnobl & Mubarakzyanov \\
$A_{4,3}$ & $A_{4.3}$ & $f_1=e_2$, $f_2=e_3$, $f_3=e_1$, $f_4=-e_4$ & $g_{4,3}$ \\
 &  & $[f_4,f_2]=f_1$, $[f_4,f_3]=f_3$ & \\
\bottomrule
\end{tabularx}
\medskip
\begin{tabulary}{\linewidth}{LLCLCLR}
\toprule
Subalgebra & Complement & dim. & Kernel & Faithful? & Shirokov realization & Popovych realization \\
\midrule
$\{0\}$ & $\langle e_1,e_2,e_3,e_4\rangle$\hphantom{xxx} & $4$ & $\{0\}$ & yes & $e_1=\partial_1$, $e_2=\partial_2$, $e_3=\partial_3$, $e_4=x_1\partial_1+x_3\partial_2+\partial_4$ & $R(A_{4.3},1)$ \\
\addlinespace
$\langle e_1\rangle$ & $\langle e_2,e_3,e_4\rangle$ & $3$ & $\langle e_1\rangle$ & no & $e_1=0$, $e_2=\partial_1$, $e_3=\partial_2$, $e_4=x_2\partial_1+\partial_3$ &  \\
\addlinespace
$\langle e_2\rangle$ & $\langle e_1,e_3,e_4\rangle$ & $3$ & $\langle e_2\rangle$ & no & $e_1=\partial_1$, $e_2=0$, $e_3=\partial_2$, $e_4=x_1\partial_1+\partial_3$ &  \\
\addlinespace
$\langle e_1+\sigma e_2\rangle$ & $\langle e_1,e_3,e_4\rangle$ & $3$ & $\{0\}$ & yes & $e_1=\partial_1$, $e_2=-\sigma e^{x_3}\partial_1$, $e_3=\partial_2$, $e_4=(x_1-\sigma x_2e^{x_3})\partial_1+\partial_3$ & $R(A_{4.3},6)$ \\
\addlinespace
$\langle e_3+\lambda e_1\rangle$ & $\langle e_1,e_2,e_4\rangle$ & $3$ & $\{0\}$ & yes & $e_1=\partial_1$, $e_2=\partial_2$, $e_3=-\lambda e^{x_3}\partial_1-x_3\partial_2$, $e_4=x_1\partial_1+\partial_3$ & $R(A_{4.3},5,\varepsilon)$ \\
\addlinespace
$\langle e_4+\lambda e_3\rangle$ & $\langle e_1,e_2,e_3\rangle$ & $3$ & $\{0\}$ & yes & $e_1=\partial_1$, $e_2=\partial_2$, $e_3=\partial_3$, $e_4=x_1\partial_1+x_3\partial_2-\lambda\partial_3$ & $R(A_{4.3},3)$ \\
\addlinespace
$\langle e_1,e_2\rangle$ & $\langle e_3,e_4\rangle$ & $2$ & $\langle e_1,e_2\rangle$ & no & $e_1=0$, $e_2=0$, $e_3=\partial_1$, $e_4=\partial_2$ &  \\
\addlinespace
$\langle e_1,e_3\rangle$ & $\langle e_2,e_4\rangle$ & $2$ & $\langle e_1\rangle$ & no & $e_1=0$, $e_2=\partial_1$, $e_3=-x_2\partial_1$, $e_4=\partial_2$ &  \\
\addlinespace
$\langle e_1+\lambda e_2,e_3\rangle$, $\lambda\neq0$ & $\langle e_1,e_4\rangle$ & $2$ & $\{0\}$ & yes & $e_1=\partial_1$, $e_2=-\lambda^{-1}e^{x_2}\partial_1$, $e_3=\lambda^{-1}x_2e^{x_2}\partial_1$, $e_4=x_1\partial_1+\partial_2$ & $R(A_{4.3},8)$ \\
\addlinespace
$\langle e_2,e_3\rangle$ & $\langle e_1,e_4\rangle$ & $2$ & $\langle e_2,e_3\rangle$ & no & $e_1=\partial_1$, $e_2=0$, $e_3=0$, $e_4=x_1\partial_1+\partial_2$ &  \\
\addlinespace
$\langle e_2,e_3+\sigma e_1\rangle$ & $\langle e_1,e_4\rangle$ & $2$ & $\langle e_2\rangle$ & no & $e_1=\partial_1$, $e_2=0$, $e_3=-\sigma e^{x_2}\partial_1$, $e_4=x_1\partial_1+\partial_2$ &  \\
\addlinespace
$\langle e_2,e_4+\lambda e_3\rangle$ & $\langle e_1,e_3\rangle$ & $2$ & $\langle e_2\rangle$ & no & $e_1=\partial_1$, $e_2=0$, $e_3=\partial_2$, $e_4=x_1\partial_1-\lambda\partial_2$ &  \\
\addlinespace
$\langle e_1,e_4+\lambda e_3\rangle$ & $\langle e_2,e_3\rangle$ & $2$ & $\langle e_1\rangle$ & no & $e_1=0$, $e_2=\partial_1$, $e_3=\partial_2$, $e_4=x_2\partial_1-\lambda\partial_2$ &  \\
\addlinespace
$\langle e_1,e_2,e_3\rangle$ & $\langle e_4\rangle$ & $1$ & $\langle e_1,e_2,e_3\rangle$ & no & $e_1=0$, $e_2=0$, $e_3=0$, $e_4=\partial_1$ &  \\
\addlinespace
$\langle e_1,e_2,e_4+\lambda e_3\rangle$ & $\langle e_3\rangle$ & $1$ & $\langle e_1,e_2,e_4+\lambda e_3\rangle$ & no & $e_1=0$, $e_2=0$, $e_3=\partial_1$, $e_4=-\lambda\partial_1$ &  \\
\addlinespace
$\langle e_2,e_3,e_4\rangle$ & $\langle e_1\rangle$ & $1$ & $\langle e_2,e_3\rangle$ & no & $e_1=\partial_1$, $e_2=0$, $e_3=0$, $e_4=x_1\partial_1$ &  \\
\bottomrule
\end{tabulary}
\caption{The algebra $s_{4,1}$}
\label{s41}
\end{table}

\subsection{The algebra $s_{4,2}$}

This algebra is the solvable four-dimensional algebra with three-dimensional Abelian nilradical and a single Jordan block of length three. 
For the rows containing the constants $\lambda^{-1}$, the comparison with Popovych uses the basis shift $\tilde e_3=e_3+\lambda^{-1}e_1$, which removes this constant term from the vector field corresponding to $e_3$. The results are summarized in table \ref{s42}.

\begin{table}[H]
\centering
\small
\begin{tabularx}{\linewidth}{lll}
\toprule
Algebra & dim. & Nonzero commutation relations \\
$s_{4,2}$ &  $4$ & $[e_1,e_4]=e_1$, $[e_2,e_4]=e_1+e_2$, $[e_3,e_4]=e_2+e_3$ \\
\bottomrule
\end{tabularx}
\begin{tabularx}{\linewidth}{llll}
Patera-Winternitz & Popovych & Šnobl & Mubarakzyanov \\
$A_{4,4}$ & $A_{4.4}$ & $f_1=e_1$, $f_2=e_2$, $f_3=e_3$, $f_4=-e_4$ & $g_{4,4}$ \\
 &  & $[f_4,f_1]=f_1$, $[f_4,f_2]=f_1+f_2$, $[f_4,f_3]=f_2+f_3$ & \\
\bottomrule
\end{tabularx}
\medskip
\begin{tabulary}{\linewidth}{LLCLCLR}
\toprule
Subalgebra & Complement & dim. & Kernel & Faithful? & Shirokov realization & Popovych realization \\
\midrule
$\{0\}$ & $\langle e_1,e_2,e_3,e_4\rangle$\hphantom{xxx} & $4$ & $\{0\}$ & yes & $e_1=\partial_1$, $e_2=\partial_2$, $e_3=\partial_3$, $e_4=(x_1+x_2)\partial_1+(x_2+x_3)\partial_2+x_3\partial_3+\partial_4$ & $R(A_{4.4},1)$ \\
\addlinespace
$\langle e_1\rangle$ & $\langle e_2,e_3,e_4\rangle$ & $3$ & $\langle e_1\rangle$ & no & $e_1=0$, $e_2=\partial_1$, $e_3=\partial_2$, $e_4=(x_1+x_2)\partial_1+x_2\partial_2+\partial_3$ &  \\
\addlinespace
$\langle e_1+\lambda e_3\rangle$, $\lambda\neq0$ & $\langle e_1,e_2,e_4\rangle$ & $3$ & $\{0\}$ & yes & $e_1=\partial_1$, $e_2=\partial_2$, $e_3=-(\tfrac12x_3^2+\lambda^{-1})\partial_1-x_3\partial_2$, $e_4=(x_1+x_2)\partial_1+x_2\partial_2+\partial_3$ & $R(A_{4.4},4)$ \\
\addlinespace
$\langle e_2\rangle$ & $\langle e_1,e_3,e_4\rangle$ & $3$ & $\{0\}$ & yes & $e_1=\partial_1$, $e_2=-x_3\partial_1$, $e_3=\partial_2$, $e_4=(x_1-x_2x_3)\partial_1+x_2\partial_2+\partial_3$ & $R(A_{4.4},5)$ \\
\addlinespace
$\langle e_3\rangle$ & $\langle e_1,e_2,e_4\rangle$ & $3$ & $\{0\}$ & yes & $e_1=\partial_1$, $e_2=\partial_2$, $e_3=-\tfrac12x_3^2\partial_1-x_3\partial_2$, $e_4=(x_1+x_2)\partial_1+x_2\partial_2+\partial_3$ & $R(A_{4.4},4)$ \\
\addlinespace
$\langle e_4\rangle$ & $\langle e_1,e_2,e_3\rangle$ & $3$ & $\{0\}$ & yes & $e_1=\partial_1$, $e_2=\partial_2$, $e_3=\partial_3$, $e_4=(x_1+x_2)\partial_1+(x_2+x_3)\partial_2+x_3\partial_3$ & $R(A_{4.4},2)$ \\
\addlinespace
$\langle e_1,e_2\rangle$ & $\langle e_3,e_4\rangle$ & $2$ & $\langle e_1,e_2\rangle$ & no & $e_1=0$, $e_2=0$, $e_3=\partial_1$, $e_4=x_1\partial_1+\partial_2$ &  \\
\addlinespace
$\langle e_1+\lambda e_3,e_2\rangle$, $\lambda\neq0$ & $\langle e_1,e_4\rangle$ & $2$ & $\{0\}$ & yes & $e_1=\partial_1$, $e_2=-x_2\partial_1$, $e_3=(\tfrac12x_2^2-\lambda^{-1})\partial_1$, $e_4=x_1\partial_1+\partial_2$ & $R(A_{4.4},7)$ \\
\addlinespace
$\langle e_1,e_3\rangle$ & $\langle e_2,e_4\rangle$ & $2$ & $\langle e_1\rangle$ & no & $e_1=0$, $e_2=\partial_1$, $e_3=-x_2\partial_1$, $e_4=x_1\partial_1+\partial_2$ &  \\
\addlinespace
$\langle e_2,e_3\rangle$ & $\langle e_1,e_4\rangle$ & $2$ & $\{0\}$ & yes & $e_1=\partial_1$, $e_2=-x_2\partial_1$, $e_3=\tfrac12x_2^2\partial_1$, $e_4=x_1\partial_1+\partial_2$ & $R(A_{4.4},7)$ \\
\addlinespace
$\langle e_1,e_4\rangle$ & $\langle e_2,e_3\rangle$ & $2$ & $\langle e_1\rangle$ & no & $e_1=0$, $e_2=\partial_1$, $e_3=\partial_2$, $e_4=(x_1+x_2)\partial_1+x_2\partial_2$ &  \\
\addlinespace
$\langle e_1,e_2,e_3\rangle$ & $\langle e_4\rangle$ & $1$ & $\langle e_1,e_2,e_3\rangle$ & no & $e_1=0$, $e_2=0$, $e_3=0$, $e_4=\partial_1$ &  \\
\addlinespace
$\langle e_1,e_2,e_4\rangle$ & $\langle e_3\rangle$ & $1$ & $\langle e_1,e_2\rangle$ & no & $e_1=0$, $e_2=0$, $e_3=\partial_1$, $e_4=x_1\partial_1$ &  \\
\bottomrule
\end{tabulary}
\caption{The algebra $s_{4,2}$}
\label{s42}
\end{table}

\subsection{The algebra $s_{4,3}(a,b)$: generic diagonal case}

This is the diagonal solvable family with three-dimensional Abelian nilradical. 
The table is written for the generic case $ab\neq0$ and the three eigenvalues $1,a,b$ pairwise different. Popovych's canonical parameter ranges are obtained from this form by permuting the three eigenvectors and rescaling the nonnilpotent generator; that is, by automorphisms of the form $\tilde e_i=e_{\pi(i)}$ for $i=1,2,3$ and $\tilde e_4=c e_4$ for some suitable $c$. The results are summarized in table \ref{s43}. 

The exceptional cases with coinciding eigenvalues are listed separately in the next sections. 

\begin{table}[H]
\centering
\small
\begin{tabularx}{\linewidth}{lll}
\toprule
Algebra &  dim. & Nonzero commutation relations \\
$s_{4,3}(a,b)$ &  $4$ & $[e_1,e_4]=e_1$, $[e_2,e_4]=a e_2$, $[e_3,e_4]=b e_3$ \\
\bottomrule
\end{tabularx}
\begin{tabulary}{\linewidth}{LLLL}
Patera-Winternitz & Popovych & Šnobl & Mubarakzyanov \\
$A_{4,5}^{a,b}$ & $A_{4.5}^{1,a,b}$ & $f_i=e_i$ for $i=1,2,3$, $f_4=-e_4$ & $g_{4,5}^{1,a,b}$ \\
 &  & $[f_4,f_1]=f_1$, $[f_4,f_2]=a f_2$, $[f_4,f_3]=b f_3$ & \\
\bottomrule
\end{tabulary}
\medskip
\begin{tabulary}{\linewidth}{LLCLCLR}
\toprule
Subalgebra & Complement & dim. & Kernel & Faithful? & Shirokov realization & Popovych realization \\
\midrule
$\{0\}$ & $\langle e_1,e_2,e_3,e_4\rangle$\hphantom{xxxx} & $4$ & $\{0\}$ & yes & $e_1=\partial_1$, $e_2=\partial_2$, $e_3=\partial_3$, $e_4=x_1\partial_1+a x_2\partial_2+b x_3\partial_3+\partial_4$ & $R(A_{4.5}^{1,a,b},1)$ \\
\addlinespace
$\langle e_1,e_2,e_3\rangle$ & $\langle e_4\rangle$ & $1$ & $\langle e_1,e_2,e_3\rangle$ & no & $e_1=0$, $e_2=0$, $e_3=0$, $e_4=\partial_1$ &  \\
\addlinespace
$\langle e_4,e_1,e_2\rangle$ & $\langle e_3\rangle$ & $1$ & $\langle e_1,e_2\rangle$ & no & $e_1=0$, $e_2=0$, $e_3=\partial_1$, $e_4=b x_1\partial_1$ &  \\
\addlinespace
$\langle e_4,e_1,e_3\rangle$ & $\langle e_2\rangle$ & $1$ & $\langle e_1,e_3\rangle$ & no & $e_1=0$, $e_2=\partial_1$, $e_3=0$, $e_4=a x_1\partial_1$ &  \\
\addlinespace
$\langle e_4,e_2,e_3\rangle$ & $\langle e_1\rangle$ & $1$ & $\langle e_2,e_3\rangle$ & no & $e_1=\partial_1$, $e_2=0$, $e_3=0$, $e_4=x_1\partial_1$ &  \\
\addlinespace
$\langle e_1,e_2\rangle$ & $\langle e_3,e_4\rangle$ & $2$ & $\langle e_1,e_2\rangle$ & no & $e_1=0$, $e_2=0$, $e_3=\partial_1$, $e_4=b x_1\partial_1+\partial_2$ &  \\
\addlinespace
$\langle e_1,e_3\rangle$ & $\langle e_2,e_4\rangle$ & $2$ & $\langle e_1,e_3\rangle$ & no & $e_1=0$, $e_2=\partial_1$, $e_3=0$, $e_4=a x_1\partial_1+\partial_2$ &  \\
\addlinespace
$\langle e_2,e_3\rangle$ & $\langle e_1,e_4\rangle$ & $2$ & $\langle e_2,e_3\rangle$ & no & $e_1=\partial_1$, $e_2=0$, $e_3=0$, $e_4=x_1\partial_1+\partial_2$ &  \\
\addlinespace
$\langle e_1,e_2+\varepsilon e_3\rangle$ & $\langle e_2,e_4\rangle$ & $2$ & $\langle e_1\rangle$ & no & $e_1=0$, $e_2=\partial_1$, $e_3=-\varepsilon e^{(a-b)x_2}\partial_1$, $e_4=a x_1\partial_1+\partial_2$ &  \\
\addlinespace
$\langle e_2,e_1+\varepsilon e_3\rangle$ & $\langle e_1,e_4\rangle$ & $2$ & $\langle e_2\rangle$ & no & $e_1=\partial_1$, $e_2=0$, $e_3=-\varepsilon e^{(1-b)x_2}\partial_1$, $e_4=x_1\partial_1+\partial_2$ &  \\
\addlinespace
$\langle e_3,e_1+\varepsilon e_2\rangle$ & $\langle e_1,e_4\rangle$ & $2$ & $\langle e_3\rangle$ & no & $e_1=\partial_1$, $e_2=-\varepsilon e^{(1-a)x_2}\partial_1$, $e_3=0$, $e_4=x_1\partial_1+\partial_2$ &  \\
\addlinespace
$\langle e_1+\varepsilon e_3,e_2+x e_3\rangle$, $x\neq0$ & $\langle e_1,e_4\rangle$ & $2$ & $\{0\}$ & yes & $e_1=\partial_1$, $e_2=\varepsilon x e^{(1-a)x_2}\partial_1$, $e_3=-\varepsilon e^{(1-b)x_2}\partial_1$, $e_4=x_1\partial_1+\partial_2$ & $R(A_{4.5}^{1,a,b},7)$ \\
\addlinespace
$\langle e_4,e_1\rangle$ & $\langle e_2,e_3\rangle$ & $2$ & $\langle e_1\rangle$ & no & $e_1=0$, $e_2=\partial_1$, $e_3=\partial_2$, $e_4=a x_1\partial_1+b x_2\partial_2$ &  \\
\addlinespace
$\langle e_4,e_2\rangle$ & $\langle e_1,e_3\rangle$ & $2$ & $\langle e_2\rangle$ & no & $e_1=\partial_1$, $e_2=0$, $e_3=\partial_2$, $e_4=x_1\partial_1+b x_2\partial_2$ &  \\
\addlinespace
$\langle e_4,e_3\rangle$ & $\langle e_1,e_2\rangle$ & $2$ & $\langle e_3\rangle$ & no & $e_1=\partial_1$, $e_2=\partial_2$, $e_3=0$, $e_4=x_1\partial_1+a x_2\partial_2$ &  \\
\addlinespace
$\langle e_1\rangle$ & $\langle e_2,e_3,e_4\rangle$ & $3$ & $\langle e_1\rangle$ & no & $e_1=0$, $e_2=\partial_1$, $e_3=\partial_2$, $e_4=a x_1\partial_1+b x_2\partial_2+\partial_3$ &  \\
\addlinespace
$\langle e_2\rangle$ & $\langle e_1,e_3,e_4\rangle$ & $3$ & $\langle e_2\rangle$ & no & $e_1=\partial_1$, $e_2=0$, $e_3=\partial_2$, $e_4=x_1\partial_1+b x_2\partial_2+\partial_3$ &  \\
\addlinespace
$\langle e_3\rangle$ & $\langle e_1,e_2,e_4\rangle$ & $3$ & $\langle e_3\rangle$ & no & $e_1=\partial_1$, $e_2=\partial_2$, $e_3=0$, $e_4=x_1\partial_1+a x_2\partial_2+\partial_3$ &  \\
\addlinespace
$\langle e_4\rangle$ & $\langle e_1,e_2,e_3\rangle$ & $3$ & $\{0\}$ & yes & $e_1=\partial_1$, $e_2=\partial_2$, $e_3=\partial_3$, $e_4=x_1\partial_1+a x_2\partial_2+b x_3\partial_3$ & $R(A_{4.5}^{1,a,b},2)$ \\
\addlinespace
$\langle e_1+\varepsilon e_3\rangle$ & $\langle e_1,e_2,e_4\rangle$ & $3$ & $\{0\}$ & yes & $e_1=\partial_1$, $e_2=\partial_2$, $e_3=-\varepsilon e^{(1-b)x_3}\partial_1$, $e_4=x_1\partial_1+a x_2\partial_2+\partial_3$ & $R(A_{4.5}^{1,a,b},5)$ \\
\addlinespace
$\langle e_2+\varepsilon e_3\rangle$ & $\langle e_1,e_2,e_4\rangle$ & $3$ & $\{0\}$ & yes & $e_1=\partial_1$, $e_2=\partial_2$, $e_3=-\varepsilon e^{(a-b)x_3}\partial_2$, $e_4=x_1\partial_1+a x_2\partial_2+\partial_3$ & $R(A_{4.5}^{1,a,b},5)$ \\
\addlinespace
$\langle e_1+\varepsilon e_2+x e_3\rangle$, $x\neq0$ & $\langle e_1,e_2,e_4\rangle$ & $3$ & $\{0\}$ & yes & $e_1=\partial_1$, $e_2=\partial_2$, $e_3=-x^{-1}e^{(1-b)x_3}\partial_1 -\varepsilon x^{-1}e^{(a-b)x_3}\partial_2$, $e_4=x_1\partial_1+a x_2\partial_2+\partial_3$ & $R(A_{4.5}^{1,a,b},5)$ \\
\bottomrule
\end{tabulary}
\caption{The algebra $s_{4,3}(a,b)$ in the generic diagonal case}
\label{s43}
\end{table}

\subsection{The special case $s_{4,3}(a,a)$}

When two or three of the eigenvalues in the diagonal family coincide, additional continuous families of subalgebras occur inside the eigenspaces. We first take the representative $s_{4,3}(a,a)$, $a \neq 0,1$.
We use a notation \begin{equation*}
    v_\phi=\cos\phi\,e_2+\sin\phi\,e_3,\quad w_\phi=-\sin\phi\,e_2+\cos\phi\,e_3.
\end{equation*} The results are summarized in table \ref{s43twoequal}.

\begin{table}[H]
\centering
\small
\begin{tabularx}{\linewidth}{lll}
\toprule
Algebra &  dim. & Nonzero commutation relations \\
$s_{4,3}(a,a)$ &  $4$ & $[e_1,e_4]=e_1$, $[e_2,e_4]=a e_2$, $[e_3,e_4]=a e_3$ \\
\bottomrule
\end{tabularx}
\begin{tabulary}{\linewidth}{LLLL}
Patera-Winternitz & Popovych & Šnobl &  Mubarakzyanov\\
$A_{4,5}^{a,a}$ & $A_{4.5}^{1,1,1/a}$  &  & $g_{4,5}^{1,1,1/a}$ \\
& $\tilde e_1=e_2$, $\tilde e_2=e_3$, $\tilde e_3=e_1$, $\tilde e_4=a^{-1}e_4$ &   $f_i=e_i$ for $i=1,2,3$, $f_4=-e_4$  & as Popovych\\
& $[\tilde e_1,\tilde e_4]=\tilde e_1$, $[\tilde e_2,\tilde e_4]=\tilde e_2$, $[\tilde e_3,\tilde e_4]=a^{-1}\tilde e_3$ & $[f_4,f_1]=f_1$, $[f_4,f_2]=a f_2$, $[f_4,f_3]=a f_3$  & \\
\bottomrule
\end{tabulary}
\medskip
\begin{tabulary}{\linewidth}{LLCLCLR}
\toprule
Subalgebra & Complement & dim. & Kernel & Faithful? & Shirokov realization & Popovych realization \\
\midrule
$\{0\}$ & $\langle e_1,e_2,e_3,e_4\rangle$\hphantom{xxx} & $4$ & $\{0\}$ & yes & $e_1=\partial_1$, $e_2=\partial_2$, $e_3=\partial_3$, $e_4=x_1\partial_1+a x_2\partial_2+a x_3\partial_3+\partial_4$ & $R(A_{4.5}^{1,1,1/a},1)$ \\
\addlinespace
$\langle e_1,e_2,e_3\rangle$ & $\langle e_4\rangle$ & $1$ & $\langle e_1,e_2,e_3\rangle$ & no & $e_1=0$, $e_2=0$, $e_3=0$, $e_4=\partial_1$ &  \\
\addlinespace
$\langle e_4,e_2,e_3\rangle$ & $\langle e_1\rangle$ & $1$ & $\langle e_2,e_3\rangle$ & no & $e_1=\partial_1$, $e_2=0$, $e_3=0$, $e_4=x_1\partial_1$ &  \\
\addlinespace
$\langle e_4,e_1,v_\phi\rangle$ & $\langle w_\phi\rangle$ & $1$ & $\langle e_1,v_\phi\rangle$ & no & $e_1=0$, $e_2=-\sin\phi\,\partial_1$, $e_3=\cos\phi\,\partial_1$, $e_4=a x_1\partial_1$ &  \\
\addlinespace
$\langle e_1,v_\phi\rangle$ & $\langle w_\phi,e_4\rangle$ & $2$ & $\langle e_1,v_\phi\rangle$ & no & $e_1=0$, $e_2=-\sin\phi\,\partial_1$, $e_3=\cos\phi\,\partial_1$, $e_4=a x_1\partial_1+\partial_2$ &  \\
\addlinespace
$\langle e_2,e_3\rangle$ & $\langle e_1,e_4\rangle$ & $2$ & $\langle e_2,e_3\rangle$ & no & $e_1=\partial_1$, $e_2=0$, $e_3=0$, $e_4=x_1\partial_1+\partial_2$ &  \\
\addlinespace
$\langle e_3,e_1+\varepsilon e_2\rangle$ & $\langle e_1,e_4\rangle$ & $2$ & $\langle e_3\rangle$ & no & $e_1=\partial_1$, $e_2=-\varepsilon e^{(1-a)x_2}\partial_1$, $e_3=0$, $e_4=x_1\partial_1+\partial_2$ &  \\
\addlinespace
$\langle e_1+\varepsilon e_3,e_2+x e_3\rangle$ & $\langle e_1,e_4\rangle$ & $2$ & $\langle e_2+x e_3\rangle$ & no & $e_1=\partial_1$, $e_2=\varepsilon x e^{(1-a)x_2}\partial_1$, $e_3=-\varepsilon e^{(1-a)x_2}\partial_1$, $e_4=x_1\partial_1+\partial_2$ &  \\
\addlinespace
$\langle e_4,e_1\rangle$ & $\langle e_2,e_3\rangle$ & $2$ & $\langle e_1\rangle$ & no & $e_1=0$, $e_2=\partial_1$, $e_3=\partial_2$, $e_4=a x_1\partial_1+a x_2\partial_2$ &  \\
\addlinespace
$\langle e_4,v_\phi\rangle$ & $\langle e_1,w_\phi\rangle$ & $2$ & $\langle v_\phi\rangle$ & no & $e_1=\partial_1$, $e_2=-\sin\phi\,\partial_2$, $e_3=\cos\phi\,\partial_2$, $e_4=x_1\partial_1+a x_2\partial_2$ &  \\
\addlinespace
$\langle e_1\rangle$ & $\langle e_2,e_3,e_4\rangle$ & $3$ & $\langle e_1\rangle$ & no & $e_1=0$, $e_2=\partial_1$, $e_3=\partial_2$, $e_4=a x_1\partial_1+a x_2\partial_2+\partial_3$ &  \\
\addlinespace
$\langle v_\phi\rangle$ & $\langle e_1,w_\phi,e_4\rangle$ & $3$ & $\langle v_\phi\rangle$ & no & $e_1=\partial_1$, $e_2=-\sin\phi\,\partial_2$, $e_3=\cos\phi\,\partial_2$, $e_4=x_1\partial_1+a x_2\partial_2+\partial_3$ &  \\
\addlinespace
$\langle e_4\rangle$ & $\langle e_1,e_2,e_3\rangle$ & $3$ & $\{0\}$ & yes & $e_1=\partial_1$, $e_2=\partial_2$, $e_3=\partial_3$, $e_4=x_1\partial_1+a x_2\partial_2+a x_3\partial_3$ & $R(A_{4.5}^{1,1,1/a},2)$ \\
\addlinespace
$\langle e_1+\varepsilon e_3\rangle$ & $\langle e_1,e_2,e_4\rangle$ & $3$ & $\{0\}$ & yes & $e_1=\partial_1$, $e_2=\partial_2$, $e_3=-\varepsilon e^{(1-a)x_3}\partial_1$, $e_4=x_1\partial_1+a x_2\partial_2+\partial_3$ & $R(A_{4.5}^{1,1,1/a},7)$ \\
\addlinespace
$\langle e_1+\varepsilon e_2\rangle$ & $\langle e_1,e_3,e_4\rangle$ & $3$ & $\{0\}$ & yes & $e_1=\partial_1$, $e_2=-\varepsilon e^{(1-a)x_3}\partial_1$, $e_3=\partial_2$, $e_4=x_1\partial_1+a x_2\partial_2+\partial_3$ & $R(A_{4.5}^{1,1,1/a},7)$ \\
\addlinespace
$\langle e_1+\varepsilon e_2+x e_3\rangle$, $x\neq0$ & $\langle e_1,e_2,e_4\rangle$ & $3$ & $\{0\}$ & yes & $e_1=\partial_1$, $e_2=\partial_2$, $e_3=-x^{-1}e^{(1-a)x_3}\partial_1-\varepsilon x^{-1}\partial_2$, $e_4=x_1\partial_1+a x_2\partial_2+\partial_3$ & $R(A_{4.5}^{1,1,1/a},7)$ \\
\bottomrule
\end{tabulary}
\caption{The exceptional diagonal algebra $s_{4,3}(a,a)$}
\label{s43twoequal}
\end{table}

\subsection{The special case $s_{4,3}(1,1)$}

For the scalar diagonal algebra all three eigenvalues coincide. If $N=\langle e_1,e_2,e_3\rangle$, every subspace $U\subset N$ is an ideal. Choosing a complement $W$ of $U$ in $N$, with basis $w_1,\ldots,w_q$ and $q=3-\dim U$, gives the compact form summarized in table \ref{s43scalar}.

\begin{table}[H]
\centering
\small
\begin{tabularx}{\linewidth}{lll}
\toprule
Algebra &  dim. & Nonzero commutation relations \\
$s_{4,3}(1,1)$ &  $4$ & $[e_i,e_4]=e_i$, $i=1,2,3$ \\
\bottomrule
\end{tabularx}
\begin{tabularx}{\linewidth}{llll}
Patera-Winternitz & Popovych & Šnobl &  Mubarakzyanov \\
 $A_{4,5}^{1,1}$ & $A_{4.5}^{1,1,1}$ & & $g_{4,5}^{1,1,1}$ \\
 & & $f_i=e_i$ for $i=1,2,3$, $f_4=-e_4$ & \\
& & $[f_4,f_1]=f_1$, $[f_4,f_2]=f_2$, $[f_4,f_3]=f_3$  & \\
\bottomrule
\end{tabularx}
\medskip
\begin{tabulary}{\linewidth}{LLCLCLR}
\toprule
Subalgebra & Complement & dim.\hphantom{xx} & Kernel & Faithful? & Shirokov realization & Popovych realization \\
\midrule
$U\subset N$ & $W\oplus\langle e_4\rangle$ & $q+1$ & $U$ & yes iff $U=\{0\}$ & $w_i=\partial_i\ (i=1,\ldots,q)$, $e_4=\sum_{i=1}^{q}x_i\partial_i+\partial_{q+1}$ & $R(A_{4.5}^{1,1,1},1)$ if $U=\{0\}$ \\
\addlinespace
$\langle e_4\rangle\oplus U$ & $W$ & $q$ & $U$ & yes iff $U=\{0\}$ & $w_i=\partial_i\ (i=1,\ldots,q)$, $e_4=\sum_{i=1}^{q}x_i\partial_i$ & $R(A_{4.5}^{1,1,1},2)$ if $U=\{0\}$ \\
\bottomrule
\end{tabulary}
\caption{The scalar diagonal algebra $s_{4,3}(1,1)$}
\label{s43scalar}
\end{table}

\subsection{The algebra $s_{4,4}(a)$}

Here $a\neq0,1$. This is the solvable algebra with a two-dimensional Jordan block and one additional real eigenvalue. 
For rows containing an arbitrary nonzero coefficient $x$ in a subalgebra representative, Popovych's normal form is obtained after rescaling the corresponding nilradical basis vector; for instance one may use $\tilde e_1=x e_1$ when the subalgebra contains $e_1+x e_2$. The results are summarized in table \ref{s44}.

The special case $a=1$ has additional Patera-Winternitz subalgebras and is treated separately.

\begin{table}[H]
\centering
\small
\begin{tabularx}{\linewidth}{lll}
\toprule
Algebra & dim. & Nonzero commutation relations \\
$s_{4,4}(a)$ & $4$ & $[e_1,e_4]=a e_1$, $[e_2,e_4]=e_2$, $[e_3,e_4]=e_2+e_3$ \\
\bottomrule
\end{tabularx}
\begin{tabularx}{\linewidth}{llll}
Patera-Winternitz & Popovych & Šnobl & Mubarakzyanov \\
$A_{4,2}^{a}$ & $A_{4.2}^{a}$ & $f_1=e_2$, $f_2=e_3$, $f_3=e_1$, $f_4=-e_4$ & $g_{4,2}^{a}$ \\
 &  & $[f_4,f_1]=f_1$, $[f_4,f_2]=f_1+f_2$, $[f_4,f_3]=a f_3$ & \\
\bottomrule
\end{tabularx}
\medskip
\begin{tabulary}{\linewidth}{LLCLCLR}
\toprule
Subalgebra & Complement & dim. & Kernel & Faithful? & Shirokov realization & Popovych realization \\
\midrule
$\{0\}$ & $\langle e_1,e_2,e_3,e_4\rangle$\hphantom{xxx} & $4$ & $\{0\}$ & yes & $e_1=\partial_1$, $e_2=\partial_2$, $e_3=\partial_3$, $e_4=a x_1\partial_1+(x_2+x_3)\partial_2+x_3\partial_3+\partial_4$ & $R(A_{4.2}^{a},1)$ \\
\addlinespace
$\langle e_1,e_2,e_3\rangle$ & $\langle e_4\rangle$ & $1$ & $\langle e_1,e_2,e_3\rangle$ & no & $e_1=0$, $e_2=0$, $e_3=0$, $e_4=\partial_1$ &  \\
\addlinespace
$\langle e_4,e_2,e_3\rangle$ & $\langle e_1\rangle$ & $1$ & $\langle e_2,e_3\rangle$ & no & $e_1=\partial_1$, $e_2=0$, $e_3=0$, $e_4=a x_1\partial_1$ &  \\
\addlinespace
$\langle e_4,e_1,e_2\rangle$ & $\langle e_3\rangle$ & $1$ & $\langle e_1,e_2\rangle$ & no & $e_1=0$, $e_2=0$, $e_3=\partial_1$, $e_4=x_1\partial_1$ &  \\
\addlinespace
$\langle e_1,e_2\rangle$ & $\langle e_3,e_4\rangle$ & $2$ & $\langle e_1,e_2\rangle$ & no & $e_1=0$, $e_2=0$, $e_3=\partial_1$, $e_4=x_1\partial_1+\partial_2$ &  \\
\addlinespace
$\langle e_1+x e_2,e_3\rangle$, $x\neq0$ & $\langle e_1,e_4\rangle$ & $2$ & $\{0\}$ & yes & $e_1=\partial_1$, $e_2=-x^{-1}e^{(a-1)x_2}\partial_1$, $e_3=x^{-1}x_2 e^{(a-1)x_2}\partial_1$, $e_4=a x_1\partial_1+\partial_2$ & $R(A_{4.2}^{a},8)$ \\
\addlinespace
$\langle e_1,e_3\rangle$ & $\langle e_2,e_4\rangle$ & $2$ & $\langle e_1\rangle$ & no & $e_1=0$, $e_2=\partial_1$, $e_3=-x_2\partial_1$, $e_4=x_1\partial_1+\partial_2$ &  \\
\addlinespace
$\langle e_2,e_3\rangle$ & $\langle e_1,e_4\rangle$ & $2$ & $\langle e_2,e_3\rangle$ & no & $e_1=\partial_1$, $e_2=0$, $e_3=0$, $e_4=a x_1\partial_1+\partial_2$ &  \\
\addlinespace
$\langle e_1+\varepsilon e_3,e_2\rangle$ & $\langle e_1,e_4\rangle$ & $2$ & $\langle e_2\rangle$ & no & $e_1=\partial_1$, $e_2=0$, $e_3=-\varepsilon e^{(a-1)x_2}\partial_1$, $e_4=a x_1\partial_1+\partial_2$ &  \\
\addlinespace
$\langle e_4,e_1\rangle$ & $\langle e_2,e_3\rangle$ & $2$ & $\langle e_1\rangle$ & no & $e_1=0$, $e_2=\partial_1$, $e_3=\partial_2$, $e_4=(x_1+x_2)\partial_1+x_2\partial_2$ &  \\
\addlinespace
$\langle e_4,e_2\rangle$ & $\langle e_1,e_3\rangle$ & $2$ & $\langle e_2\rangle$ & no & $e_1=\partial_1$, $e_2=0$, $e_3=\partial_2$, $e_4=a x_1\partial_1+x_2\partial_2$ &  \\
\addlinespace
$\langle e_1\rangle$ & $\langle e_2,e_3,e_4\rangle$ & $3$ & $\langle e_1\rangle$ & no & $e_1=0$, $e_2=\partial_1$, $e_3=\partial_2$, $e_4=(x_1+x_2)\partial_1+x_2\partial_2+\partial_3$ &  \\
\multicolumn{7}{c}{(continued on next page)}\end{tabulary}\end{table}\begin{table}[H]\centering\small\begin{tabulary}{\linewidth}{LLCLCLR} Subalgebra & Complement & dim. & Kernel & Faithful? & Shirokov realization & Popovych realization \\ \midrule \multicolumn{7}{c}{(continued from previous page)} \\
$\langle e_2\rangle$ & $\langle e_1,e_3,e_4\rangle$ & $3$ & $\langle e_2\rangle$ & no & $e_1=\partial_1$, $e_2=0$, $e_3=\partial_2$, $e_4=a x_1\partial_1+x_2\partial_2+\partial_3$ &  \\
\addlinespace
$\langle e_4\rangle$ & $\langle e_1,e_2,e_3\rangle$ & $3$ & $\{0\}$ & yes & $e_1=\partial_1$, $e_2=\partial_2$, $e_3=\partial_3$, $e_4=a x_1\partial_1+(x_2+x_3)\partial_2+x_3\partial_3$ & $R(A_{4.2}^{a},2)$ \\
\addlinespace
$\langle e_1+\varepsilon e_2\rangle$ & $\langle e_1,e_3,e_4\rangle$ & $3$ & $\{0\}$ & yes & $e_1=\partial_1$, $e_2=-\varepsilon e^{(a-1)x_3}\partial_1$, $e_3=\partial_2$, $e_4=(a x_1-\varepsilon x_2e^{(a-1)x_3})\partial_1+x_2\partial_2+\partial_3$ & $R(A_{4.2}^{a},7)$ \\
\addlinespace
$\langle e_3+x e_1\rangle$ & $\langle e_1,e_2,e_4\rangle$ & $3$ & $\{0\}$ & yes & $e_1=\partial_1$, $e_2=\partial_2$, $e_3=-x e^{(a-1)x_3}\partial_1-x_3\partial_2$, $e_4=a x_1\partial_1+x_2\partial_2+\partial_3$ & $R(A_{4.2}^{a},7)$ \\
\bottomrule
\end{tabulary}
\caption{The algebra $s_{4,4}(a)$}
\label{s44}
\end{table}

\subsection{The exceptional case $s_{4,4}(1)$}

Here we put $u_\phi=\cos\phi\,e_1+\sin\phi\,e_2$ and $w_\phi=-\sin\phi\,e_1+\cos\phi\,e_2$. In the row $\langle e_3+x e_1\rangle$, the automorphic basis change $\tilde e_3=e_3+x e_1$ gives Popovych's realization. The results are summarized in table \ref{s441}.

\begin{table}[H]
\centering
\small
\begin{tabularx}{\linewidth}{lll}
\toprule
Algebra &  dim. & Nonzero commutation relations \\
$s_{4,4}(1)$ &  $4$ & $[e_1,e_4]=e_1$, $[e_2,e_4]=e_2$, $[e_3,e_4]=e_2+e_3$ \\
\bottomrule
\end{tabularx}
\begin{tabularx}{\linewidth}{llll}
Patera-Winternitz & Popovych & Šnobl &  Mubarakzyanov \\
 $A_{4,2}^{1}$ & $A_{4.2}^{1}$ & $f_1=e_2$, $f_2=e_3$, $f_3=e_1$, $f_4=-e_4$  & $g_{4,2}^{1}$ \\
 & & $[f_4,f_1]=f_1$, $[f_4,f_2]=f_1+f_2$, $[f_4,f_3]= f_3$  & \\
\bottomrule
\end{tabularx}
\medskip
\begin{tabulary}{\linewidth}{LLCLCLR}
\toprule
Subalgebra & Complement & dim. & Kernel & Faithful? & Shirokov realization & Popovych realization \\
\midrule
$\{0\}$ & $\langle e_1,e_2,e_3,e_4\rangle$\hphantom{xxx} & $4$ & $\{0\}$ & yes & $e_1=\partial_1$, $e_2=\partial_2$, $e_3=\partial_3$, $e_4=x_1\partial_1+(x_2+x_3)\partial_2+x_3\partial_3+\partial_4$ & $R(A_{4.2}^{1},1)$ \\
\addlinespace
$\langle e_1,e_2,e_3\rangle$ & $\langle e_4\rangle$ & $1$ & $\langle e_1,e_2,e_3\rangle$ & no & $e_1=0$, $e_2=0$, $e_3=0$, $e_4=\partial_1$ &  \\
\addlinespace
$\langle e_4,e_2,e_3+x e_1\rangle$ & $\langle e_1\rangle$ & $1$ & $\langle e_2,e_3+x e_1\rangle$ & no & $e_1=\partial_1$, $e_2=0$, $e_3=-x\partial_1$, $e_4=x_1\partial_1$ &  \\
\addlinespace
$\langle e_4,e_1,e_2\rangle$ & $\langle e_3\rangle$ & $1$ & $\langle e_1,e_2\rangle$ & no & $e_1=0$, $e_2=0$, $e_3=\partial_1$, $e_4=x_1\partial_1$ &  \\
\addlinespace
$\langle e_1,e_2\rangle$ & $\langle e_3,e_4\rangle$ & $2$ & $\langle e_1,e_2\rangle$ & no & $e_1=0$, $e_2=0$, $e_3=\partial_1$, $e_4=x_1\partial_1+\partial_2$ &  \\
\addlinespace
$\langle e_1+x e_2,e_3\rangle$ & $\langle e_2,e_4\rangle$ & $2$ & $\langle e_1+x e_2\rangle$ & no & $e_1=-x\partial_1$, $e_2=\partial_1$, $e_3=-x_2\partial_1$, $e_4=x_1\partial_1+\partial_2$ &  \\
\addlinespace
$\langle e_2,e_3+x e_1\rangle$ & $\langle e_1,e_4\rangle$ & $2$ & $\langle e_2,e_3+x e_1\rangle$ & no & $e_1=\partial_1$, $e_2=0$, $e_3=-x\partial_1$, $e_4=x_1\partial_1+\partial_2$ &  \\
\addlinespace
$\langle e_4,u_\phi\rangle$ & $\langle w_\phi,e_3\rangle$ & $2$ & $\langle u_\phi\rangle$ & no & $e_1=-\sin\phi\,\partial_1$, $e_2=\cos\phi\,\partial_1$, $e_3=\partial_2$, $e_4=(x_1+x_2\cos\phi)\partial_1+x_2\partial_2$ &  \\
\addlinespace
$\langle u_\phi\rangle$ & $\langle w_\phi,e_3,e_4\rangle$ & $3$ & $\langle u_\phi\rangle$ & no & $e_1=-\sin\phi\,\partial_1$, $e_2=\cos\phi\,\partial_1$, $e_3=\partial_2$, $e_4=(x_1+x_2\cos\phi)\partial_1+x_2\partial_2+\partial_3$ &  \\
\addlinespace
$\langle e_3+x e_1\rangle$ & $\langle e_1,e_2,e_4\rangle$ & $3$ & $\{0\}$ & yes & $e_1=\partial_1$, $e_2=\partial_2$, $e_3=-x\partial_1-x_3\partial_2$, $e_4=x_1\partial_1+x_2\partial_2+\partial_3$ & $R(A_{4.2}^{1},4)$ \\
\addlinespace
$\langle e_4\rangle$ & $\langle e_1,e_2,e_3\rangle$ & $3$ & $\{0\}$ & yes & $e_1=\partial_1$, $e_2=\partial_2$, $e_3=\partial_3$, $e_4=x_1\partial_1+(x_2+x_3)\partial_2+x_3\partial_3$ & $R(A_{4.2}^{1},2)$ \\
\bottomrule
\end{tabulary}
\caption{The exceptional algebra $s_{4,4}(1)$}
\label{s441}
\end{table}

\subsection{The algebra $s_{4,5}(a,b)$}

This is the solvable family with three-dimensional Abelian nilradical for which the action of the nonnilpotent generator has one real eigenvalue and one complex conjugate pair.
The entries with $x>0$ are the nonzero part of the Patera-Winternitz family $\langle e_1+x e_3\rangle$ and its two-dimensional analogue; the endpoint $x=0$ is represented separately by the rows involving $\langle e_1\rangle$ or $\langle e_1,e_2\rangle$. The comparison with Popovych uses the local coordinate changes
\begin{equation*}
    y_2=\tan x_2\quad\text{or}\quad y_3=\tan x_3,
\end{equation*}
according to which quotient coordinate occurs in the row. The constant $x>0$ is removed by rescaling the nilradical vector in the corresponding subalgebra representative. The results are summarized in table \ref{s45}.

\begin{table}[H]
\centering
\small
\begin{tabularx}{\linewidth}{lll}
\toprule
Algebra & dim. & Nonzero commutation relations \\
$s_{4,5}(a,b)$ & $4$ & $[e_1,e_4]=a e_1$, $[e_2,e_4]=b e_2-e_3$, $[e_3,e_4]=e_2+b e_3$ \\
\bottomrule
\end{tabularx}
\begin{tabularx}{\linewidth}{llll}
Patera-Winternitz & Popovych & Šnobl & Mubarakzyanov \\
$A_{4,6}^{a,b}$ & $A_{4.6}^{a,b}$ & $f_i=e_i$ for $i=1,2,3$, $f_4=-e_4$ & $g_{4,6}^{a,b}$ \\
 &  & $[f_4,f_1]=a f_1$, $[f_4,f_2]=b f_2-f_3$, $[f_4,f_3]=f_2+b f_3$ & \\
\bottomrule
\end{tabularx}
\medskip
\begin{tabulary}{\linewidth}{LLCLCLR}
\toprule
Subalgebra & Complement & dim. & Kernel & Faithful? & Shirokov realization & Popovych realization \\
\midrule
$\{0\}$ & $\langle e_1,e_2,e_3,e_4\rangle$\hphantom{xxxxxxx} & $4$ & $\{0\}$ & yes & $e_1=\partial_1$, $e_2=\partial_2$, $e_3=\partial_3$, $e_4=a x_1\partial_1+(b x_2+x_3)\partial_2+(b x_3-x_2)\partial_3+\partial_4$ & $R(A_{4.6}^{a,b},1)$ \\
\addlinespace
$\langle e_1,e_2,e_3\rangle$ & $\langle e_4\rangle$ & $1$ & $\langle e_1,e_2,e_3\rangle$\hphantom{xxxx} & no & $e_1=0$, $e_2=0$, $e_3=0$, $e_4=\partial_1$ &  \\
\addlinespace
$\langle e_4,e_2,e_3\rangle$ & $\langle e_1\rangle$ & $1$ & $\langle e_2,e_3\rangle$ & no & $e_1=\partial_1$, $e_2=0$, $e_3=0$, $e_4=a x_1\partial_1$ &  \\
\addlinespace
$\langle e_1,e_2\rangle$ & $\langle e_3,e_4\rangle$ & $2$ & $\langle e_1\rangle$ & no & $e_1=0$, $e_2=\tan x_2\,\partial_1$, $e_3=\partial_1$, $e_4=x_1(b+\tan x_2)\partial_1+\partial_2$ &  \\
\addlinespace
$\langle e_1+x e_3,e_2\rangle$, $x>0$ & $\langle e_1,e_4\rangle$ & $2$ & $\{0\}$ & yes & $e_1=\partial_1$, $e_2=-x^{-1}e^{(a-b)x_2}\sin x_2\,\partial_1$, $e_3=-x^{-1}e^{(a-b)x_2}\cos x_2\,\partial_1$, $e_4=a x_1\partial_1+\partial_2$ & $R(A_{4.6}^{a,b},6)$ \\
\addlinespace
$\langle e_2,e_3\rangle$ & $\langle e_1,e_4\rangle$ & $2$ & $\langle e_2,e_3\rangle$ & no & $e_1=\partial_1$, $e_2=0$, $e_3=0$, $e_4=a x_1\partial_1+\partial_2$ &  \\
\addlinespace
$\langle e_4,e_1\rangle$ & $\langle e_2,e_3\rangle$ & $2$ & $\langle e_1\rangle$ & no & $e_1=0$, $e_2=\partial_1$, $e_3=\partial_2$, $e_4=(b x_1+x_2)\partial_1+(b x_2-x_1)\partial_2$ &  \\
\addlinespace
$\langle e_1\rangle$ & $\langle e_2,e_3,e_4\rangle$ & $3$ & $\langle e_1\rangle$ & no & $e_1=0$, $e_2=\partial_1$, $e_3=\partial_2$, $e_4=(b x_1+x_2)\partial_1+(b x_2-x_1)\partial_2+\partial_3$ &  \\
\addlinespace
$\langle e_1+x e_3\rangle$, $x>0$ & $\langle e_1,e_2,e_4\rangle$ & $3$ & $\{0\}$ & yes & $e_1=\partial_1$, $e_2=\partial_2$, $e_3=-x^{-1}e^{(a-b)x_3}\sec x_3\,\partial_1-\tan x_3\,\partial_2$, $e_4=(a x_1+x^{-1}x_2e^{(a-b)x_3}\sec x_3)\partial_1 +x_2(b+\tan x_3)\partial_2+\partial_3$ & $R(A_{4.6}^{a,b},4)$ \\
\addlinespace
$\langle e_3\rangle$ & $\langle e_1,e_2,e_4\rangle$ & $3$ & $\{0\}$ & yes & $e_1=\partial_1$, $e_2=\partial_2$, $e_3=-\tan x_3\,\partial_2$, $e_4=a x_1\partial_1+x_2(b+\tan x_3)\partial_2+\partial_3$ & $R(A_{4.6}^{a,b},4)$ \\
\addlinespace
$\langle e_4\rangle$ & $\langle e_1,e_2,e_3\rangle$ & $3$ & $\{0\}$ & yes & $e_1=\partial_1$, $e_2=\partial_2$, $e_3=\partial_3$, $e_4=a x_1\partial_1+(b x_2+x_3)\partial_2+(b x_3-x_2)\partial_3$ & $R(A_{4.6}^{a,b},2)$ \\
\bottomrule
\end{tabulary}
\caption{The algebra $s_{4,5}(a,b)$}
\label{s45}
\end{table}

\subsection{The algebra $s_{4,6}$}

The algebra $s_{4,6}$ is a solvable algebra with Heisenberg nilradical $n_{3,1}$ and split derivation. 
The rows with $e_4+x e_1$ are compared with Popovych after the central automorphism
\begin{equation*}
    \tilde e_1=e_1,\qquad \tilde e_2=e_2,\qquad \tilde e_3=e_3,\qquad \tilde e_4=e_4+x e_1.
\end{equation*}
For the row $\langle e_2\rangle$ we additionally use $\tilde e_1=-e_1$, $\tilde e_2=e_3$, $\tilde e_3=e_2$, $\tilde e_4=-e_4$ and $y_3=-x_3$. For the row $\langle e_2+\varepsilon e_3\rangle$ we use $y_1=x_1$, $y_2=x_2$, $y_3=-\varepsilon e^{2x_3}$. These changes put the faithful rows into Popovych's forms. The results are summarized in table \ref{s46}.

\begin{table}[H]
\centering
\small
\begin{tabularx}{\linewidth}{lll}
\toprule
Algebra &  dim. & Nonzero commutation relations \\
$s_{4,6}$ &  $4$ & $[e_2,e_3]=e_1$, $[e_2,e_4]=e_2$, $[e_3,e_4]=-e_3$ \\
\bottomrule
\end{tabularx}
\begin{tabularx}{\linewidth}{llll}
Patera-Winternitz & Popovych & Šnobl & Mubarakzyanov \\
$A_{4,8}$ & $A_{4.8}^{-1}$ & $f_i=e_i$ for $i=1,2,3$, $f_4=-e_4$ & $g_{4,8}^{-1}$ \\
 &  & $[f_2,f_3]=f_1$, $[f_4,f_2]=f_2$, $[f_4,f_3]=-f_3$ & \\
\bottomrule
\end{tabularx}
\medskip
\begin{tabulary}{\linewidth}{LLCLCLR}
\toprule
Subalgebra & Complement & dim. & Kernel & Faithful? & Shirokov realization & Popovych realization \\
\midrule
$\{0\}$ & $\langle e_1,e_2,e_3,e_4\rangle$ & $4$ & $\{0\}$ & yes & $e_1=\partial_1$, $e_2=\partial_2$, $e_3=x_2\partial_1+\partial_3$, $e_4=x_2\partial_2-x_3\partial_3+\partial_4$ & $R(A_{4.8}^{-1},1)$ \\
\addlinespace
$\langle e_1,e_2,e_3\rangle$ & $\langle e_4\rangle$ & $1$ & $\langle e_1,e_2,e_3\rangle$ & no & $e_1=0$, $e_2=0$, $e_3=0$, $e_4=\partial_1$ &  \\
\addlinespace
$\langle e_4,e_1,e_2\rangle$ & $\langle e_3\rangle$ & $1$ & $\langle e_1,e_2\rangle$ & no & $e_1=0$, $e_2=0$, $e_3=\partial_1$, $e_4=-x_1\partial_1$ &  \\
\addlinespace
$\langle e_4,e_1,e_3\rangle$ & $\langle e_2\rangle$ & $1$ & $\langle e_1,e_3\rangle$ & no & $e_1=0$, $e_2=\partial_1$, $e_3=0$, $e_4=x_1\partial_1$ &  \\
\addlinespace
$\langle e_1,e_2\rangle$ & $\langle e_3,e_4\rangle$ & $2$ & $\langle e_1,e_2\rangle$ & no & $e_1=0$, $e_2=0$, $e_3=\partial_1$, $e_4=-x_1\partial_1+\partial_2$ &  \\
\addlinespace
$\langle e_1,e_3\rangle$ & $\langle e_2,e_4\rangle$ & $2$ & $\langle e_1,e_3\rangle$ & no & $e_1=0$, $e_2=\partial_1$, $e_3=0$, $e_4=x_1\partial_1+\partial_2$ &  \\
\addlinespace
$\langle e_1,e_2+\varepsilon e_3\rangle$ & $\langle e_2,e_4\rangle$ & $2$ & $\langle e_1\rangle$ & no & $e_1=0$, $e_2=\partial_1$, $e_3=-\varepsilon e^{2x_2}\partial_1$, $e_4=x_1\partial_1+\partial_2$ &  \\
\addlinespace
$\langle e_4,e_1\rangle$ & $\langle e_2,e_3\rangle$ & $2$ & $\langle e_1\rangle$ & no & $e_1=0$, $e_2=\partial_1$, $e_3=\partial_2$, $e_4=x_1\partial_1-x_2\partial_2$ &  \\
\addlinespace
$\langle e_4+x e_1,e_2\rangle$ & $\langle e_1,e_3\rangle$ & $2$ & $\{0\}$ & yes & $e_1=\partial_1$, $e_2=-x_2\partial_1$, $e_3=\partial_2$, $e_4=-x\partial_1-x_2\partial_2$ & $R(A_{4.8}^{-1},5)$ \\
\addlinespace
$\langle e_4+x e_1,e_3\rangle$ & $\langle e_1,e_2\rangle$ & $2$ & $\{0\}$ & yes & $e_1=\partial_1$, $e_2=\partial_2$, $e_3=x_2\partial_1$, $e_4=-x\partial_1+x_2\partial_2$ & $R(A_{4.8}^{-1},5)$ \\
\addlinespace
$\langle e_1\rangle$ & $\langle e_2,e_3,e_4\rangle$ & $3$ & $\langle e_1\rangle$ & no & $e_1=0$, $e_2=\partial_1$, $e_3=\partial_2$, $e_4=x_1\partial_1-x_2\partial_2+\partial_3$ &  \\
\addlinespace
$\langle e_2\rangle$ & $\langle e_1,e_3,e_4\rangle$ & $3$ & $\{0\}$ & yes & $e_1=\partial_1$, $e_2=-x_2\partial_1$, $e_3=\partial_2$, $e_4=-x_2\partial_2+\partial_3$ & $R(A_{4.8}^{-1},4)$ \\
\addlinespace
$\langle e_2+\varepsilon e_3\rangle$ & $\langle e_1,e_2,e_4\rangle$ & $3$ & $\{0\}$ & yes & $e_1=\partial_1$, $e_2=\partial_2$, $e_3=x_2\partial_1-\varepsilon e^{2x_3}\partial_2$, $e_4=x_2\partial_2+\partial_3$ & $R(A_{4.8}^{-1},3)$ \\
\addlinespace
$\langle e_3\rangle$ & $\langle e_1,e_2,e_4\rangle$ & $3$ & $\{0\}$ & yes & $e_1=\partial_1$, $e_2=\partial_2$, $e_3=x_2\partial_1$, $e_4=x_2\partial_2+\partial_3$ & $R(A_{4.8}^{-1},4)$ \\
\addlinespace
$\langle e_4+x e_1\rangle$ & $\langle e_1,e_2,e_3\rangle$ & $3$ & $\{0\}$ & yes & $e_1=\partial_1$, $e_2=\partial_2$, $e_3=x_2\partial_1+\partial_3$, $e_4=-x\partial_1+x_2\partial_2-x_3\partial_3$ & $R(A_{4.8}^{-1},2)$ \\
\bottomrule
\end{tabulary}
\caption{The algebra $s_{4,6}$}
\label{s46}
\end{table}

\subsection{The algebra $s_{4,7}$}

This is the four-dimensional oscillator algebra, also often called the Nappi-Witten algebra. It has Heisenberg nilradical and a rotational derivation. 
For rows containing $e_4+x e_1$ we use the central automorphism
\begin{equation*}
    \tilde e_1=e_1,\qquad \tilde e_2=e_2,\qquad \tilde e_3=e_3,\qquad \tilde e_4=e_4+x e_1.
\end{equation*}
For the row $\langle e_2\rangle$ the comparison with Popovych is obtained with $y_1=x_1$, $y_2=x_2$, $y_3=-\tan x_3$. The results are summarized in table \ref{s47}.

\begin{table}[H]
\centering
\small
\begin{tabularx}{\linewidth}{llll}
\toprule
Algebra & Also & dim. & Nonzero commutation relations \\
$s_{4,7}$ & oscillator/Nappi-Witten & $4$ & $[e_2,e_3]=e_1$, $[e_2,e_4]=-e_3$, $[e_3,e_4]=e_2$ \\
\bottomrule
\end{tabularx}
\begin{tabularx}{\linewidth}{llll}
Patera-Winternitz & Popovych & Šnobl & Mubarakzyanov \\
$A_{4,10}$ & $A_{4.9}^{0}$ & $f_i=e_i$ for $i=1,2,3$, $f_4=-e_4$ & $g_{4,9}^{0}$ \\
 &  & $[f_2,f_3]=f_1$, $[f_4,f_2]=-f_3$, $[f_4,f_3]=f_2$ & \\
\bottomrule
\end{tabularx}
\medskip
\begin{tabulary}{\linewidth}{LLCLCLR}
\toprule
Subalgebra & Complement & dim. & Kernel & Faithful? & Shirokov realization & Popovych realization \\
\midrule
$\{0\}$ & $\langle e_1,e_2,e_3,e_4\rangle$\hphantom{xxx} & $4$ & $\{0\}$ & yes & $e_1=\partial_1$, $e_2=\partial_2$, $e_3=x_2\partial_1+\partial_3$, $e_4=\frac{x_3^2-x_2^2}{2}\partial_1+x_3\partial_2-x_2\partial_3+\partial_4$ & $R(A_{4.9}^{0},1)$ \\
\addlinespace
$\langle e_1,e_2,e_3\rangle$ & $\langle e_4\rangle$ & $1$ & $\langle e_1,e_2,e_3\rangle$ & no & $e_1=0$, $e_2=0$, $e_3=0$, $e_4=\partial_1$ &  \\
\addlinespace
$\langle e_1,e_2\rangle$ & $\langle e_3,e_4\rangle$ & $2$ & $\langle e_1\rangle$ & no & $e_1=0$, $e_2=\tan x_2\partial_1$, $e_3=\partial_1$, $e_4=x_1\tan x_2\partial_1+\partial_2$ &  \\
\addlinespace
$\langle e_1,e_4\rangle$ & $\langle e_2,e_3\rangle$ & $2$ & $\langle e_1\rangle$ & no & $e_1=0$, $e_2=\partial_1$, $e_3=\partial_2$, $e_4=x_2\partial_1-x_1\partial_2$ &  \\
\addlinespace
$\langle e_1\rangle$ & $\langle e_2,e_3,e_4\rangle$ & $3$ & $\langle e_1\rangle$ & no & $e_1=0$, $e_2=\partial_1$, $e_3=\partial_2$, $e_4=x_2\partial_1-x_1\partial_2+\partial_3$ &  \\
\addlinespace
$\langle e_2\rangle$ & $\langle e_1,e_3,e_4\rangle$ & $3$ & $\{0\}$ & yes & $e_1=\partial_1$, $e_2=-x_2\partial_1+\tan x_3\partial_2$, $e_3=\partial_2$, $e_4=-\frac{x_2^2}{2}\partial_1+x_2\tan x_3\partial_2+\partial_3$ & $R(A_{4.9}^{0},3)$ \\
\addlinespace
$\langle e_4+x e_1\rangle$ & $\langle e_1,e_2,e_3\rangle$ & $3$ & $\{0\}$ & yes & $e_1=\partial_1$, $e_2=\partial_2$, $e_3=x_2\partial_1+\partial_3$, $e_4=\left(-x-\frac{x_2^2}{2}+\frac{x_3^2}{2}\right)\partial_1+x_3\partial_2-x_2\partial_3$ & $R(A_{4.9}^{0},2)$ \\
\bottomrule
\end{tabulary}
\caption{The algebra $s_{4,7}$}
\label{s47}
\end{table}

\subsection{The algebra $s_{4,8}(a)$: generic case}

Here $-1<a<1$ and $a\neq0$. This is the Heisenberg-nilradical family. 
The endpoint $a=-1$ is $s_{4,6}$ and the case $a=0$ is $s_{4,11}$. For the rows $\langle e_4,e_2\rangle$ and $\langle e_4,e_3\rangle$ the comparison with Popovych only uses the sign/permutation automorphisms
\begin{equation*}
    (\tilde e_1,\tilde e_2,\tilde e_3,\tilde e_4)=(e_1,-e_2,e_3,e_4)
    \quad\text{or}\quad
    (e_1,e_3,e_2,e_4),
\end{equation*}
as appropriate. For the row $\langle e_2+\varepsilon e_3\rangle$ one may use
\begin{equation*}
    y_1=x_1,\qquad y_2=x_2,\qquad y_3=-\varepsilon e^{(1-a)x_3}.
\end{equation*}
The results are summarized in table \ref{s48}.

\begin{table}[H]
\centering
\small
\begin{tabularx}{\linewidth}{llll}
\toprule
Algebra & Also & dim. & Nonzero commutation relations \\
$s_{4,8}(a)$ & Heisenberg nilradical & $4$ & $[e_2,e_3]=e_1$, $[e_1,e_4]=(1+a)e_1$, $[e_2,e_4]=e_2$, $[e_3,e_4]=a e_3$ \\
\bottomrule
\end{tabularx}
\begin{tabulary}{\linewidth}{LLLL}
Patera-Winternitz & Popovych & Šnobl & Mubarakzyanov \\
$A_{4,9}^{a}$ & $A_{4.8}^{a}$ & $f_i=e_i$ for $i=1,2,3$, $f_4=-e_4$ & $g_{4,8}^{a}$ \\
 &  & $[f_2,f_3]=f_1$, $[f_4,f_1]=(1+a)f_1$, $[f_4,f_2]=f_2$, $[f_4,f_3]=a f_3$ & \\
\bottomrule
\end{tabulary}
\medskip
\begin{tabulary}{\linewidth}{LLCLCLR}
\toprule
Subalgebra & Complement & dim. & Kernel & Faithful? & Shirokov realization & Popovych realization \\
\midrule
$\{0\}$ & $\langle e_1,e_2,e_3,e_4\rangle$\hphantom{xxx} & $4$ & $\{0\}$ & yes & $e_1=\partial_1$, $e_2=\partial_2$, $e_3=x_2\partial_1+\partial_3$, $e_4=(1+a)x_1\partial_1+x_2\partial_2+a x_3\partial_3+\partial_4$ & $R(A_{4.8}^{a},1)$ \\
\addlinespace
$\langle e_1,e_2,e_3\rangle$ & $\langle e_4\rangle$ & $1$ & $\langle e_1,e_2,e_3\rangle$ & no & $e_1=0$, $e_2=0$, $e_3=0$, $e_4=\partial_1$ &  \\
\addlinespace
$\langle e_4,e_1,e_2\rangle$ & $\langle e_3\rangle$ & $1$ & $\langle e_1,e_2\rangle$ & no & $e_1=0$, $e_2=0$, $e_3=\partial_1$, $e_4=a x_1\partial_1$ &  \\
\addlinespace
$\langle e_4,e_1,e_3\rangle$ & $\langle e_2\rangle$ & $1$ & $\langle e_1,e_3\rangle$ & no & $e_1=0$, $e_2=\partial_1$, $e_3=0$, $e_4=x_1\partial_1$ &  \\
\addlinespace
$\langle e_1,e_2\rangle$ & $\langle e_3,e_4\rangle$ & $2$ & $\langle e_1,e_2\rangle$ & no & $e_1=0$, $e_2=0$, $e_3=\partial_1$, $e_4=a x_1\partial_1+\partial_2$ &  \\
\addlinespace
$\langle e_1,e_3\rangle$ & $\langle e_2,e_4\rangle$ & $2$ & $\langle e_1,e_3\rangle$ & no & $e_1=0$, $e_2=\partial_1$, $e_3=0$, $e_4=x_1\partial_1+\partial_2$ &  \\
\multicolumn{7}{c}{(continued on next page)}\end{tabulary}\end{table}\begin{table}[H]\centering\small\begin{tabulary}{\linewidth}{LLCLCLR} Subalgebra & Complement & dim. & Kernel & Faithful? & Shirokov realization & Popovych realization \\ \midrule \multicolumn{7}{c}{(continued from previous page)} \\
$\langle e_1,e_2+\varepsilon e_3\rangle$ & $\langle e_2,e_4\rangle$ & $2$ & $\langle e_1\rangle$ & no & $e_1=0$, $e_2=\partial_1$, $e_3=-\varepsilon e^{(1-a)x_2}\partial_1$, $e_4=x_1\partial_1+\partial_2$ &  \\
\addlinespace
$\langle e_4,e_1\rangle$ & $\langle e_2,e_3\rangle$ & $2$ & $\langle e_1\rangle$ & no & $e_1=0$, $e_2=\partial_1$, $e_3=\partial_2$, $e_4=x_1\partial_1+a x_2\partial_2$ &  \\
\addlinespace
$\langle e_4,e_2\rangle$ & $\langle e_1,e_3\rangle$ & $2$ & $\{0\}$ & yes & $e_1=\partial_1$, $e_2=-x_2\partial_1$, $e_3=\partial_2$, $e_4=(1+a)x_1\partial_1+a x_2\partial_2$ & $R(A_{4.8}^{a},7)$ \\
\addlinespace
$\langle e_4,e_3\rangle$ & $\langle e_1,e_2\rangle$ & $2$ & $\{0\}$ & yes & $e_1=\partial_1$, $e_2=\partial_2$, $e_3=x_2\partial_1$, $e_4=(1+a)x_1\partial_1+x_2\partial_2$ & $R(A_{4.8}^{a},5)$ \\
\addlinespace
$\langle e_1\rangle$ & $\langle e_2,e_3,e_4\rangle$ & $3$ & $\langle e_1\rangle$ & no & $e_1=0$, $e_2=\partial_1$, $e_3=\partial_2$, $e_4=x_1\partial_1+a x_2\partial_2+\partial_3$ &  \\
\addlinespace
$\langle e_2\rangle$ & $\langle e_1,e_3,e_4\rangle$ & $3$ & $\{0\}$ & yes & $e_1=\partial_1$, $e_2=-x_2\partial_1$, $e_3=\partial_2$, $e_4=(1+a)x_1\partial_1+a x_2\partial_2+\partial_3$ & $R(A_{4.8}^{a},6)$ \\
\addlinespace
$\langle e_3\rangle$ & $\langle e_1,e_2,e_4\rangle$ & $3$ & $\{0\}$ & yes & $e_1=\partial_1$, $e_2=\partial_2$, $e_3=x_2\partial_1$, $e_4=(1+a)x_1\partial_1+x_2\partial_2+\partial_3$ & $R(A_{4.8}^{a},4)$ \\
\addlinespace
$\langle e_4\rangle$ & $\langle e_1,e_2,e_3\rangle$ & $3$ & $\{0\}$ & yes & $e_1=\partial_1$, $e_2=\partial_2$, $e_3=x_2\partial_1+\partial_3$, $e_4=(1+a)x_1\partial_1+x_2\partial_2+a x_3\partial_3$ & $R(A_{4.8}^{a},2)$ \\
\addlinespace
$\langle e_2+\varepsilon e_3\rangle$ & $\langle e_1,e_2,e_4\rangle$ & $3$ & $\{0\}$ & yes & $e_1=\partial_1$, $e_2=\partial_2$, $e_3=x_2\partial_1-\varepsilon e^{(1-a)x_3}\partial_2$, $e_4=(1+a)x_1\partial_1+x_2\partial_2+\partial_3$ & $R(A_{4.8}^{a},3)$ \\
\bottomrule
\end{tabulary}
\caption{The algebra $s_{4,8}(a)$, generic case}
\label{s48}
\end{table}

\subsection{The special case $s_{4,8}(1)$}

For $a=1$ in the previous family,
Patera-Winternitz list additional representatives involving the line $\langle e_2\cos\phi+e_3\sin\phi\rangle$. We put $v_\phi=e_2\cos\phi+e_3\sin\phi$ and $w_\phi=-e_2\sin\phi+e_3\cos\phi$. The results are summarized in table \ref{s48endpoint}.

\begin{table}[H]
\centering
\small
\begin{tabularx}{\linewidth}{lll}
\toprule
Algebra &  dim. & Nonzero commutation relations \\
$s_{4,8}(1)$ &  $4$ & $[e_2,e_3]=e_1$, $[e_1,e_4]=2e_1$, $[e_2,e_4]=e_2$, $[e_3,e_4]=e_3$ \\
\bottomrule
\end{tabularx}
\begin{tabularx}{\linewidth}{llll}
Patera-Winternitz & Popovych & Šnobl & Mubarakzyanov \\
$A_{4,9}^{1}$ & $A_{4.8}^{1}$ &  $f_i=e_i$ for $i=1,2,3$, $f_4=-e_4$ & $g_{4,8}^{1}$ \\
 &  & $[f_2,f_3]=f_1$, $[f_4,f_1]=2f_1$, $[f_4,f_2]=f_2$, $[f_4,f_3]=f_3$ & \\
\bottomrule
\end{tabularx}
\medskip
\begin{tabulary}{\linewidth}{LLCLCLR}
\toprule
Subalgebra & Complement & dim. & Kernel & Faithful? & Shirokov realization & Popovych realization \\
\midrule
$\{0\}$ & $\langle e_1,e_2,e_3,e_4\rangle$\hphantom{xxxx} & $4$ & $\{0\}$ & yes & $e_1=\partial_1$, $e_2=\partial_2$, $e_3=x_2\partial_1+\partial_3$, $e_4=2x_1\partial_1+x_2\partial_2+x_3\partial_3+\partial_4$ & $R(A_{4.8}^{1},1)$ \\
\addlinespace
$\langle e_1,e_2,e_3\rangle$\hphantom{xxx} & $\langle e_4\rangle$ & $1$ & $\langle e_1,e_2,e_3\rangle$\hphantom{x} & no & $e_1=0$, $e_2=0$, $e_3=0$, $e_4=\partial_1$ &  \\
\addlinespace
$\langle e_4,e_1,v_\phi\rangle$ & $\langle w_\phi\rangle$ & $1$ & $\langle e_1,v_\phi\rangle$ & no & $e_1=0$, $e_2=-\sin\phi\,\partial_1$, $e_3=\cos\phi\,\partial_1$, $e_4=x_1\partial_1$ &  \\
\addlinespace
$\langle e_1,v_\phi\rangle$ & $\langle w_\phi,e_4\rangle$ & $2$ & $\langle e_1,v_\phi\rangle$ & no & $e_1=0$, $e_2=-\sin\phi\,\partial_1$, $e_3=\cos\phi\,\partial_1$, $e_4=x_1\partial_1+\partial_2$ &  \\
\addlinespace
$\langle e_4,e_1\rangle$ & $\langle e_2,e_3\rangle$ & $2$ & $\langle e_1\rangle$ & no & $e_1=0$, $e_2=\partial_1$, $e_3=\partial_2$, $e_4=x_1\partial_1+x_2\partial_2$ &  \\
\addlinespace
$\langle e_4,v_\phi\rangle$ & $\langle e_1,w_\phi\rangle$ & $2$ & $\{0\}$ & yes & $e_1=\partial_1$, $e_2=-x_2\cos\phi\,\partial_1-\sin\phi\,\partial_2$, $e_3=-x_2\sin\phi\,\partial_1+\cos\phi\,\partial_2$, $e_4=2x_1\partial_1+x_2\partial_2$ & $R(A_{4.8}^{1},6)$ \\
\addlinespace
$\langle e_1\rangle$ & $\langle e_2,e_3,e_4\rangle$ & $3$ & $\langle e_1\rangle$ & no & $e_1=0$, $e_2=\partial_1$, $e_3=\partial_2$, $e_4=x_1\partial_1+x_2\partial_2+\partial_3$ &  \\
\addlinespace
$\langle v_\phi\rangle$ & $\langle e_1,w_\phi,e_4\rangle$ & $3$ & $\{0\}$ & yes & $e_1=\partial_1$, $e_2=-x_2\cos\phi\,\partial_1-\sin\phi\,\partial_2$, $e_3=-x_2\sin\phi\,\partial_1+\cos\phi\,\partial_2$, $e_4=2x_1\partial_1+x_2\partial_2+\partial_3$ & $R(A_{4.8}^{1},6)$ \\
\addlinespace
$\langle e_4\rangle$ & $\langle e_1,e_2,e_3\rangle$ & $3$ & $\{0\}$ & yes & $e_1=\partial_1$, $e_2=\partial_2$, $e_3=x_2\partial_1+\partial_3$, $e_4=2x_1\partial_1+x_2\partial_2+x_3\partial_3$ & $R(A_{4.8}^{1},2)$ \\
\bottomrule
\end{tabulary}
\caption{The special case $s_{4,8}(1)$}
\label{s48endpoint}
\end{table}

\subsection{The algebra $s_{4,9}(\alpha)$}

Here $\alpha>0$. This Heisenberg-nilradical algebra has a rotational derivation together with a dilation.
For the row $\langle e_2\rangle$, one comparison with Popovych's printed normal form is obtained by
\begin{equation*}
    y_1=x_1,\qquad y_2=x_2,\qquad y_3=-\tan x_3,
\end{equation*}
and by the basis change $\tilde e_1=e_1$, $\tilde e_2=e_3$, $\tilde e_3=-e_2$, $\tilde e_4=e_4$. The results are summarized in table \ref{s49}.

\begin{table}[H]
\centering
\small
\begin{tabularx}{\linewidth}{lll}
\toprule
Algebra &  dim. & Nonzero commutation relations \\
$s_{4,9}(\alpha)$ &  $4$ & $[e_2,e_3]=e_1$, $[e_1,e_4]=2\alpha e_1$, $[e_2,e_4]=\alpha e_2-e_3$, $[e_3,e_4]=e_2+\alpha e_3$ \\
\bottomrule
\end{tabularx}
\begin{tabulary}{\linewidth}{LLLL}
Patera-Winternitz & Popovych & Šnobl & Mubarakzyanov \\
$A_{4,11}^{\alpha}$ & $A_{4.9}^{\alpha}$ & $f_i=e_i$ for $i=1,2,3$, $f_4=-e_4$ & $g_{4,9}^{\alpha}$ \\
 & & $[f_2,f_3]=f_1$, $[f_4,f_1]=2\alpha f_1$, $[f_4,f_2]=\alpha f_2-f_3$, $[f_4,f_3]=f_2+\alpha f_3$ & \\
\bottomrule
\end{tabulary}
\medskip
\begin{tabulary}{\linewidth}{LLCLCLR}
\toprule
Subalgebra & Complement & dim. & Kernel & Faithful? & Shirokov realization & Popovych realization \\
\midrule
$\{0\}$ & $\langle e_1,e_2,e_3,e_4\rangle$\hphantom{xxxxx} & $4$ & $\{0\}$ & yes & $e_1=\partial_1$, $e_2=\partial_2$, $e_3=x_2\partial_1+\partial_3$, $e_4=\left(2\alpha x_1-\frac{x_2^2}{2}+\frac{x_3^2}{2}\right)\partial_1 +(\alpha x_2+x_3)\partial_2+(\alpha x_3-x_2)\partial_3+\partial_4$ & $R(A_{4.9}^{\alpha},1)$ \\
\addlinespace
$\langle e_1,e_2,e_3\rangle$\hphantom{xxx} & $\langle e_4\rangle$ & $1$ & $\langle e_1,e_2,e_3\rangle$\hphantom{xx} & no & $e_1=0$, $e_2=0$, $e_3=0$, $e_4=\partial_1$ &  \\
\addlinespace
$\langle e_1,e_2\rangle$ & $\langle e_3,e_4\rangle$ & $2$ & $\langle e_1\rangle$ & no & $e_1=0$, $e_2=\tan x_2 \partial_1$, $e_3=\partial_1$, $e_4=x_1(\alpha+\tan x_2)\partial_1+\partial_2$ &  \\
\addlinespace
$\langle e_4,e_1\rangle$ & $\langle e_2,e_3\rangle$ & $2$ & $\langle e_1\rangle$ & no & $e_1=0$, $e_2=\partial_1$, $e_3=\partial_2$, $e_4=(\alpha x_1+x_2)\partial_1+(\alpha x_2-x_1)\partial_2$ &  \\
\addlinespace
$\langle e_1\rangle$ & $\langle e_2,e_3,e_4\rangle$ & $3$ & $\langle e_1\rangle$ & no & $e_1=0$, $e_2=\partial_1$, $e_3=\partial_2$, $e_4=(\alpha x_1+x_2)\partial_1+(\alpha x_2-x_1)\partial_2+\partial_3$ &  \\
\addlinespace
$\langle e_2\rangle$ & $\langle e_1,e_3,e_4\rangle$ & $3$ & $\{0\}$ & yes & $e_1=\partial_1$, $e_2=-x_2\partial_1+\tan x_3 \partial_2$, $e_3=\partial_2$, $e_4=\left(2\alpha x_1-\frac{x_2^2}{2}\right)\partial_1 +x_2(\alpha+\tan x_3)\partial_2+\partial_3$ & $R(A_{4.9}^{\alpha},3)$ \\
\addlinespace
$\langle e_4\rangle$ & $\langle e_1,e_2,e_3\rangle$ & $3$ & $\{0\}$ & yes & $e_1=\partial_1$, $e_2=\partial_2$, $e_3=x_2\partial_1+\partial_3$, $e_4=\left(2\alpha x_1-\frac{x_2^2}{2}+\frac{x_3^2}{2}\right)\partial_1 +(\alpha x_2+x_3)\partial_2+(\alpha x_3-x_2)\partial_3$ & $R(A_{4.9}^{\alpha},2)$ \\
\bottomrule
\end{tabulary}
\caption{The algebra $s_{4,9}(\alpha)$}
\label{s49}
\end{table}

\subsection{The algebra $s_{4,10}$}

This is a special Heisenberg-nilradical algebra with a nonsemisimple derivation. 
In the row $\langle e_3\rangle$, the coordinate substitution $y_3=-x_3$ gives Popovych's printed form. In the rows $\langle e_2\rangle$ and $\langle e_4,e_2\rangle$, changing the sign of the second coordinate gives the printed forms. The results are summarized in table \ref{s410}.

\begin{table}[H]
\centering
\small
\begin{tabularx}{\linewidth}{lll}
\toprule
Algebra &  dim. & Nonzero commutation relations \\
$s_{4,10}$ &  $4$ & $[e_2,e_3]=e_1$, $[e_1,e_4]=2e_1$, $[e_2,e_4]=e_2$, $[e_3,e_4]=e_2+e_3$ \\
\bottomrule
\end{tabularx}
\begin{tabulary}{\linewidth}{LLLL}
Patera-Winternitz & Popovych & Šnobl & Mubarakzyanov \\
$A_{4,7}$ & $A_{4.7}$ & $f_i=e_i$ for $i=1,2,3$, $f_4=-e_4$ & $g_{4,7}$ \\
& & $[f_2,f_3]=f_1$, $[f_4,f_1]=2f_1$, $[f_4,f_2]=f_2$, $[f_4,f_3]=f_2+f_3$ & \\
\bottomrule
\end{tabulary}
\medskip
\begin{tabulary}{\linewidth}{LLCLCLR}
\toprule
Subalgebra & Complement & dim. & Kernel & Faithful? & Shirokov realization & Popovych realization \\
\midrule
$\{0\}$ & $\langle e_1,e_2,e_3,e_4\rangle$\hphantom{xxx} & $4$ & $\{0\}$ & yes & $e_1=\partial_1$, $e_2=\partial_2$, $e_3=x_2\partial_1+\partial_3$, $e_4=\left(2x_1+\frac{x_3^2}{2}\right)\partial_1+(x_2+x_3)\partial_2+x_3\partial_3+\partial_4$ & $R(A_{4.7},1)$ \\
\addlinespace
$\langle e_1,e_2,e_3\rangle$\hphantom{xx} & $\langle e_4\rangle$ & $1$ & $\langle e_1,e_2,e_3\rangle$\hphantom{x} & no & $e_1=0$, $e_2=0$, $e_3=0$, $e_4=\partial_1$ &  \\
\addlinespace
$\langle e_4,e_1,e_2\rangle$ & $\langle e_3\rangle$ & $1$ & $\langle e_1,e_2\rangle$ & no & $e_1=0$, $e_2=0$, $e_3=\partial_1$, $e_4=x_1\partial_1$ &  \\
\multicolumn{7}{c}{(continued on next page)}\end{tabulary}\end{table}\begin{table}[H]\centering\small\begin{tabulary}{\linewidth}{LLCLCLR} Subalgebra & Complement & dim. & Kernel & Faithful? & Shirokov realization & Popovych realization \\ \midrule \multicolumn{7}{c}{(continued from previous page)} \\
$\langle e_1,e_2\rangle$ & $\langle e_3,e_4\rangle$ & $2$ & $\langle e_1,e_2\rangle$ & no & $e_1=0$, $e_2=0$, $e_3=\partial_1$, $e_4=x_1\partial_1+\partial_2$ &  \\
\addlinespace
$\langle e_1,e_3\rangle$ & $\langle e_2,e_4\rangle$ & $2$ & $\langle e_1\rangle$ & no & $e_1=0$, $e_2=\partial_1$, $e_3=-x_2\partial_1$, $e_4=x_1\partial_1+\partial_2$ &  \\
\addlinespace
$\langle e_4,e_1\rangle$ & $\langle e_2,e_3\rangle$ & $2$ & $\langle e_1\rangle$ & no & $e_1=0$, $e_2=\partial_1$, $e_3=\partial_2$, $e_4=(x_1+x_2)\partial_1+x_2\partial_2$ &  \\
\addlinespace
$\langle e_4,e_2\rangle$ & $\langle e_1,e_3\rangle$ & $2$ & $\{0\}$ & yes & $e_1=\partial_1$, $e_2=-x_2\partial_1$, $e_3=\partial_2$, $e_4=\left(2x_1-\frac{x_2^2}{2}\right)\partial_1+x_2\partial_2$ & $R(A_{4.7},5)$ \\
\addlinespace
$\langle e_1\rangle$ & $\langle e_2,e_3,e_4\rangle$ & $3$ & $\langle e_1\rangle$ & no & $e_1=0$, $e_2=\partial_1$, $e_3=\partial_2$, $e_4=(x_1+x_2)\partial_1+x_2\partial_2+\partial_3$ &  \\
\addlinespace
$\langle e_2\rangle$ & $\langle e_1,e_3,e_4\rangle$ & $3$ & $\{0\}$ & yes & $e_1=\partial_1$, $e_2=-x_2\partial_1$, $e_3=\partial_2$, $e_4=\left(2x_1-\frac{x_2^2}{2}\right)\partial_1+x_2\partial_2+\partial_3$ & $R(A_{4.7},4)$ \\
\addlinespace
$\langle e_3\rangle$ & $\langle e_1,e_2,e_4\rangle$ & $3$ & $\{0\}$ & yes & $e_1=\partial_1$, $e_2=\partial_2$, $e_3=x_2\partial_1-x_3\partial_2$, $e_4=2x_1\partial_1+x_2\partial_2+\partial_3$ & $R(A_{4.7},3)$ \\
\addlinespace
$\langle e_4\rangle$ & $\langle e_1,e_2,e_3\rangle$ & $3$ & $\{0\}$ & yes & $e_1=\partial_1$, $e_2=\partial_2$, $e_3=x_2\partial_1+\partial_3$, $e_4=\left(2x_1+\frac{x_3^2}{2}\right)\partial_1+(x_2+x_3)\partial_2+x_3\partial_3$ & $R(A_{4.7},2)$ \\
\bottomrule
\end{tabulary}
\caption{The algebra $s_{4,10}$}
\label{s410}
\end{table}

\subsection{The algebra $s_{4,11}$}

This is another special case of the Heisenberg-nilradical family. 
In the row $\langle e_2+\varepsilon e_3\rangle$, the change of variables
\begin{equation*}
    y_1=x_1,\qquad y_2=x_2,\qquad y_3=-\varepsilon e^{x_3}
\end{equation*}
gives Popovych's printed form $R(A_{4.8}^{0},3)$. The results are summarized in table \ref{s411}.

\begin{table}[H]
\centering
\small
\begin{tabularx}{\linewidth}{lll}
\toprule
Algebra &  dim. & Nonzero commutation relations \\
$s_{4,11}$ &  $4$ & $[e_2,e_3]=e_1$, $[e_1,e_4]=e_1$, $[e_2,e_4]=e_2$ \\
\bottomrule
\end{tabularx}
\begin{tabularx}{\linewidth}{llll}
Patera-Winternitz & Popovych & Šnobl & Mubarakzyanov \\
$A_{4,9}^{0}$ & $A_{4.8}^{0}$ & $f_i=e_i$ for $i=1,2,3$, $f_4=-e_4$ & $g_{4,8}^{0}$ \\
 & & $[f_2,f_3]=f_1$, $[f_4,f_1]=f_1$, $[f_4,f_2]=f_2$ & \\
\bottomrule
\end{tabularx}
\medskip
\begin{tabulary}{\linewidth}{LLCLCLR}
\toprule
Subalgebra & Complement & dim. & Kernel & Faithful? & Shirokov realization & Popovych realization \\
\midrule
$\{0\}$ & $\langle e_1,e_2,e_3,e_4\rangle$\hphantom{xx} & $4$ & $\{0\}$ & yes & $e_1=\partial_1$, $e_2=\partial_2$, $e_3=x_2\partial_1+\partial_3$, $e_4=x_1\partial_1+x_2\partial_2+\partial_4$ & $R(A_{4.8}^{0},1)$ \\
\addlinespace
$\langle e_1,e_2,e_3\rangle$ & $\langle e_4\rangle$ & $1$ & $\langle e_1,e_2,e_3\rangle$ & no & $e_1=0$, $e_2=0$, $e_3=0$, $e_4=\partial_1$ &  \\
\addlinespace
$\langle e_3,e_4,e_1\rangle$ & $\langle e_2\rangle$ & $1$ & $\langle e_1,e_3\rangle$ & no & $e_1=0$, $e_2=\partial_1$, $e_3=0$, $e_4=x_1\partial_1$ &  \\
\addlinespace
$\langle e_4,e_1,e_2\rangle$ & $\langle e_3\rangle$ & $1$ & $\langle e_1,e_2,e_4\rangle$ & no & $e_1=0$, $e_2=0$, $e_3=\partial_1$, $e_4=0$ &  \\
\addlinespace
$\langle e_4+x e_3,e_1,e_2\rangle$ & $\langle e_3\rangle$ & $1$ & $\langle e_1,e_2,e_4+x e_3\rangle$ & no & $e_1=0$, $e_2=0$, $e_3=\partial_1$, $e_4=-x\partial_1$ &  \\
\addlinespace
$\langle e_1,e_2\rangle$ & $\langle e_3,e_4\rangle$ & $2$ & $\langle e_1,e_2\rangle$ & no & $e_1=0$, $e_2=0$, $e_3=\partial_1$, $e_4=\partial_2$ &  \\
\addlinespace
$\langle e_1,e_3\rangle$ & $\langle e_2,e_4\rangle$ & $2$ & $\langle e_1,e_3\rangle$ & no & $e_1=0$, $e_2=\partial_1$, $e_3=0$, $e_4=x_1\partial_1+\partial_2$ &  \\
\addlinespace
$\langle e_1,e_2+\varepsilon e_3\rangle$ & $\langle e_2,e_4\rangle$ & $2$ & $\langle e_1\rangle$ & no & $e_1=0$, $e_2=\partial_1$, $e_3=-\varepsilon e^{x_2}\partial_1$, $e_4=x_1\partial_1+\partial_2$ &  \\
\addlinespace
$\langle e_3,e_4\rangle$ & $\langle e_1,e_2\rangle$ & $2$ & $\{0\}$ & yes & $e_1=\partial_1$, $e_2=\partial_2$, $e_3=x_2\partial_1$, $e_4=x_1\partial_1+x_2\partial_2$ & $R(A_{4.8}^{0},5)$ \\
\addlinespace
$\langle e_4+x e_3,e_1\rangle$ & $\langle e_2,e_3\rangle$ & $2$ & $\langle e_1\rangle$ & no & $e_1=0$, $e_2=\partial_1$, $e_3=\partial_2$, $e_4=x_1\partial_1-x\partial_2$ &  \\
\multicolumn{7}{c}{(continued on next page)}\end{tabulary}\end{table}\begin{table}[H]\centering\small\begin{tabulary}{\linewidth}{LLCLCLR} Subalgebra & Complement & dim. & Kernel & Faithful? & Shirokov realization & Popovych realization \\ \midrule \multicolumn{7}{c}{(continued from previous page)} \\
$\langle e_4,e_2\rangle$ & $\langle e_1,e_3\rangle$ & $2$ & $\{0\}$ & yes & $e_1=\partial_1$, $e_2=-x_2\partial_1$, $e_3=\partial_2$, $e_4=x_1\partial_1$ & $R(A_{4.8}^{0},7)$ \\
\addlinespace
$\langle e_1\rangle$ & $\langle e_2,e_3,e_4\rangle$ & $3$ & $\langle e_1\rangle$ & no & $e_1=0$, $e_2=\partial_1$, $e_3=\partial_2$, $e_4=x_1\partial_1+\partial_3$ &  \\
\addlinespace
$\langle e_2\rangle$ & $\langle e_1,e_3,e_4\rangle$ & $3$ & $\{0\}$ & yes & $e_1=\partial_1$, $e_2=-x_2\partial_1$, $e_3=\partial_2$, $e_4=x_1\partial_1+\partial_3$ & $R(A_{4.8}^{0},6)$ \\
\addlinespace
$\langle e_3\rangle$ & $\langle e_1,e_2,e_4\rangle$ & $3$ & $\{0\}$ & yes & $e_1=\partial_1$, $e_2=\partial_2$, $e_3=x_2\partial_1$, $e_4=x_1\partial_1+x_2\partial_2+\partial_3$ & $R(A_{4.8}^{0},4)$ \\
\addlinespace
$\langle e_2+\varepsilon e_3\rangle$ & $\langle e_1,e_2,e_4\rangle$ & $3$ & $\{0\}$ & yes & $e_1=\partial_1$, $e_2=\partial_2$, $e_3=x_2\partial_1-\varepsilon e^{x_3}\partial_2$, $e_4=x_1\partial_1+x_2\partial_2+\partial_3$ & $R(A_{4.8}^{0},3)$ \\
\addlinespace
$\langle e_4+x e_3\rangle$ & $\langle e_1,e_2,e_3\rangle$ & $3$ & $\{0\}$ & yes & $e_1=\partial_1$, $e_2=\partial_2$, $e_3=x_2\partial_1+\partial_3$, $e_4=x_1\partial_1+x_2\partial_2-x\partial_3$ & $R(A_{4.8}^{0},9,C)$, $C=-x$ \\
\bottomrule
\end{tabulary}
\caption{The algebra $s_{4,11}$}
\label{s411}
\end{table}

\subsection{The algebra $s_{4,12}$}

This is the Lie algebra of the orientation-preserving similitude group of the Euclidean plane, often denoted $\mathrm{sim}(2)$. 
For the rows $\langle e_1\rangle$ and $\langle e_3,e_1\rangle$, the substitutions
\begin{equation*}
    y_2=\tan x_3\qquad\hbox{and respectively}\qquad y_2=\tan x_2
\end{equation*}
together with the basis change $\tilde e_1=e_2$, $\tilde e_2=e_1$, $\tilde e_3=e_3$, $\tilde e_4=-e_4$ put the realizations into Popovych's printed forms. The results are summarized in table \ref{s412}.

\begin{table}[H]
\centering
\small
\begin{tabularx}{\linewidth}{llll}
\toprule
Algebra & Also & dim. & Nonzero commutation relations \\
$s_{4,12}$ & $\mathrm{sim}(2)$ & $4$ & $[e_1,e_3]=e_1$, $[e_2,e_3]=e_2$, $[e_1,e_4]=-e_2$, $[e_2,e_4]=e_1$ \\
\bottomrule
\end{tabularx}
\begin{tabularx}{\linewidth}{llll}
Patera-Winternitz & Popovych & Šnobl & Mubarakzyanov \\
$A_{4,12}$ & $A_{4.10}$ & $f_1=e_1$, $f_2=e_2$, $f_3=-e_3$, $f_4=-e_4$ & $g_{4,10}$ \\
& & $[f_3,f_1]=f_1$, $[f_3,f_2]=f_2$, $[f_4,f_1]=-f_2$, $[f_4,f_2]=f_1$ & \\
\bottomrule
\end{tabularx}
\medskip
\begin{tabulary}{\linewidth}{LLCLCLR}
\toprule
Subalgebra & Complement & dim. & Kernel & Faithful? & Shirokov realization & Popovych realization \\
\midrule
$\{0\}$ & $\langle e_1,e_2,e_3,e_4\rangle$\hphantom{xxx} & $4$ & $\{0\}$ & yes & $e_1=\partial_1$, $e_2=\partial_2$, $e_3=x_1\partial_1+x_2\partial_2+\partial_3$, $e_4=x_2\partial_1-x_1\partial_2+\partial_4$ & $R(A_{4.10},1)$ \\
\addlinespace
$\langle e_3,e_1,e_2\rangle$ & $\langle e_4\rangle$ & $1$ & $\langle e_1,e_2,e_3\rangle$ & no & $e_1=0$, $e_2=0$, $e_3=0$, $e_4=\partial_1$ &  \\
\addlinespace
$\langle e_4,e_1,e_2\rangle$ & $\langle e_3\rangle$ & $1$ & $\langle e_1,e_2,e_4\rangle$ & no & $e_1=0$, $e_2=0$, $e_3=\partial_1$, $e_4=0$ &  \\
\addlinespace
$\langle e_4+x e_3,e_1,e_2\rangle$ & $\langle e_3\rangle$ & $1$ & $\langle e_1,e_2,e_4+x e_3\rangle$ & no & $e_1=0$, $e_2=0$, $e_3=\partial_1$, $e_4=-x\partial_1$ &  \\
\addlinespace
$\langle e_1,e_2\rangle$ & $\langle e_3,e_4\rangle$ & $2$ & $\langle e_1,e_2\rangle$ & no & $e_1=0$, $e_2=0$, $e_3=\partial_1$, $e_4=\partial_2$ &  \\
\addlinespace
$\langle e_3,e_4\rangle$ & $\langle e_1,e_2\rangle$ & $2$ & $\{0\}$ & yes & $e_1=\partial_1$, $e_2=\partial_2$, $e_3=x_1\partial_1+x_2\partial_2$, $e_4=x_2\partial_1-x_1\partial_2$ & $R(A_{4.10},6)$ \\
\addlinespace
$\langle e_3,e_1\rangle$ & $\langle e_2,e_4\rangle$ & $2$ & $\{0\}$ & yes & $e_1=\tan x_2 \partial_1$, $e_2=\partial_1$, $e_3=x_1\partial_1$, $e_4=x_1\tan x_2 \partial_1+\partial_2$ & $R(A_{4.10},7)$ \\
\addlinespace
$\langle e_1\rangle$ & $\langle e_2,e_3,e_4\rangle$ & $3$ & $\{0\}$ & yes & $e_1=\tan x_3 \partial_1$, $e_2=\partial_1$, $e_3=x_1\partial_1+\partial_2$, $e_4=x_1\tan x_3 \partial_1+\partial_3$ & $R(A_{4.10},4)$ \\
\addlinespace
$\langle e_3\rangle$ & $\langle e_1,e_2,e_4\rangle$ & $3$ & $\{0\}$ & yes & $e_1=\partial_1$, $e_2=\partial_2$, $e_3=x_1\partial_1+x_2\partial_2$, $e_4=x_2\partial_1-x_1\partial_2+\partial_3$ & $R(A_{4.10},5)$ \\
\addlinespace
$\langle e_4+x e_3\rangle$ & $\langle e_1,e_2,e_3\rangle$ & $3$ & $\{0\}$ & yes & $e_1=\partial_1$, $e_2=\partial_2$, $e_3=x_1\partial_1+x_2\partial_2+\partial_3$, $e_4=x_2\partial_1-x_1\partial_2-x\partial_3$ & $R(A_{4.10},3,C)$, $C=-x$ \\
\bottomrule
\end{tabulary}
\caption{The algebra $s_{4,12}$}
\label{s412}
\end{table}

\section{Conclusion}
We have computed the homogeneous realizations obtained directly from Shirokov's construction for the low-dimensional real Lie algebras up to dimension four, using the Patera-Winternitz representatives of subalgebras. This gives, for each isotropy subalgebra, the corresponding projected realization, its kernel, and hence whether the associated homogeneous action is effective. The faithful effective rows were then compared with well-known Popovych's weak-equivalence classification of faithful realizations.

To wrap things up, it’s worth taking a look at some statistics (see table \ref{summarytable}). 
The last column of the table counts faithful rows for which a Popovych realization is explicitly identified. The clean total for the finite, explicitly enumerated indecomposable tables is 244 Shirokov rows,\footnote{In table \ref{s43scalar}, the scalar diagonal algebra $s_{4,3}(1,1)$ is written in a compact schematic form. The two displayed rows contain faithful specializations when $U=\{0\}$, giving the two Popovych rows recorded in the table. For nonzero subspaces $U\subset N$, the corresponding realizations are nonfaithful and factor through quotient algebras. Therefore this table is not included in the numerical total.} of which 106 are faithful, 138 unfaithful, and 106 Popovych rows were recovered.

\begin{table}[H]
\centering
\small
\begin{tabulary}{\linewidth}{LCRRRR}
\toprule
Algebra & Table & Shirokov rows & Faithful rows & Nonfaithful rows & Popovych rows recovered \\
\midrule
$n_{1,1}$ & \ref{n11} & $1$ & $1$ & $0$ & $1$ \\
$s_{2,1}$ & \ref{s21} & $3$ & $2$ & $1$ & $2$ \\
$n_{3,1}$ & \ref{n31} & $4$ & $2$ & $2$ & $2$ \\
$s_{3,1}(a)$, $a\neq1$ & \ref{s31generic} & $8$ & $3$ & $5$ & $3$ \\
$s_{3,1}(1)$ & \ref{s31special} & $5$ & $2$ & $3$ & $2$ \\
$s_{3,2}$ & \ref{s32} & $6$ & $3$ & $3$ & $3$ \\
$s_{3,3}(a)$ & \ref{s33} & $4$ & $3$ & $1$ & $3$ \\
$\mathrm{sl}(2,\mathbb R)$ & \ref{sl2r} & $5$ & $5$ & $0$ & $5$ \\
$\mathrm{so}(3)$ & \ref{so3} & $2$ & $2$ & $0$ & $2$ \\
\midrule
$n_{4,1}$ & \ref{n41} & $11$ & $5$ & $6$ & $5$ \\
$s_{4,1}$ & \ref{s41} & $16$ & $5$ & $11$ & $5$ \\
$s_{4,2}$ & \ref{s42} & $13$ & $7$ & $6$ & $7$ \\
$s_{4,3}(a,b)$ & \ref{s43} & $22$ & $6$ & $16$ & $6$ \\
$s_{4,3}(a,a)$ & \ref{s43twoequal} & $16$ & $5$ & $11$ & $5$ \\
$s_{4,3}(1,1)$ & \ref{s43scalar} & $2$ & $2$ & see note\footnotemark[1] & $2$ \\
$s_{4,4}(a)$ & \ref{s44} & $16$ & $5$ & $11$ & $5$ \\
$s_{4,4}(1)$ & \ref{s441} & $11$ & $3$ & $8$ & $3$ \\
$s_{4,5}(a,b)$ & \ref{s45} & $11$ & $5$ & $6$ & $5$ \\
$s_{4,6}$ & \ref{s46} & $15$ & $7$ & $8$ & $7$ \\
$s_{4,7}$ & \ref{s47} & $7$ & $3$ & $4$ & $3$ \\
$s_{4,8}(a)$ & \ref{s48} & $15$ & $7$ & $8$ & $7$ \\
$s_{4,8}(1)$ & \ref{s48endpoint} & $9$ & $4$ & $5$ & $4$ \\
$s_{4,9}(\alpha)$ & \ref{s49} & $7$ & $3$ & $4$ & $3$ \\
$s_{4,10}$ & \ref{s410} & $11$ & $5$ & $6$ & $5$ \\
$s_{4,11}$ & \ref{s411} & $16$ & $7$ & $9$ & $7$ \\
$s_{4,12}$ & \ref{s412} & $10$ & $6$ & $4$ & $6$ \\
\midrule
Total, excluding case \ref{s43scalar}
&  & $244$ & $106$ & $138$ & $106$ \\
\bottomrule
\end{tabulary}
\caption{Summary of the Shirokov realization tables.}
\label{summarytable}
\end{table}

\begin{appendices}
\section{Decomposable algebras}
\label{decomposable}

The preceding sections contain the indecomposable real Lie algebras of dimension at most four. Let us comment here the decomposable cases only in a brief, compact form. Šnobl's notation $n_{1,1}$ is used for the one-dimensional Abelian algebra and $s_{2,1}$ for the two-dimensional non-Abelian algebra with $[e_1,e_2]=e_1$ (see tables \ref{n11} and \ref{s21}).

We use the following natural convention. If
\begin{equation*}
    g=g_1\oplus g_2
\end{equation*}
is a direct product of Lie algebras $g_1$ and $g_2$ and if $R_1$ and $R_2$ are Shirokov realizations of $g_1$ and $g_2$ in disjoint variables, then the direct product realization of $g$ is obtained by putting the vector fields of $R_1$ and $R_2$ side by side. In particular, the generic Shirokov realization of a decomposable algebra is the direct product of the generic Shirokov realizations of its indecomposable summands.

As is well known, there exist Goursat-type, or diagonal, subalgebras of direct sums. These may give additional homogeneous realizations, and they are responsible for the fact that several Shirokov rows can correspond to the same Popovych row after weak equivalence. In this appendix we record the generic rows and the simplest direct-product rows, and we give the algebra $s_{2,1}\oplus n_{1,1}$ as a model example of how a Goursat-type row can be displayed.

The decomposable real Lie algebras of dimension at most four are listed in table \ref{decomposablelist}. 
The notation in terms of the algebras used in the present paper is given in the first column, while the Popovych notation is used in the second column.

\begin{table}[H]
\centering
\small
\begin{tabulary}{\linewidth}{CLLL}
\toprule
Dim. & Notation used here & Popovych notation & Nonzero commutation relations \\
\midrule
$2$ & $2n_{1,1}$ & $2A_1$ & none \\
\addlinespace
$3$ & $3n_{1,1}$ & $3A_1$ & none \\
$3$ & $s_{2,1}\oplus n_{1,1}$ & $A_{2.1}\oplus A_1$ & $[e_1,e_2]=e_1$ \\
\addlinespace
$4$ & $4n_{1,1}$ & $4A_1$ & none \\
$4$ & $s_{2,1}\oplus 2n_{1,1}$ & $A_{2.1}\oplus 2A_1$ & $[e_1,e_2]=e_1$ \\
$4$ & $2s_{2,1}$ & $2A_{2.1}$ & $[e_1,e_2]=e_1$, $[e_3,e_4]=e_3$ \\
$4$ & $n_{3,1}\oplus n_{1,1}$ & $A_{3.1}\oplus A_1$ & $[e_2,e_3]=e_1$ \\
$4$ & $s_{3,2}\oplus n_{1,1}$ & $A_{3.2}\oplus A_1$ & $[e_1,e_3]=e_1$, $[e_2,e_3]=e_1+e_2$ \\
$4$ & $s_{3,1}(1)\oplus n_{1,1}$ & $A_{3.3}\oplus A_1$ & $[e_1,e_3]=e_1$, $[e_2,e_3]=e_2$ \\
$4$ & $s_{3,1}(a)\oplus n_{1,1}$ & $A^a_{3.4}\oplus A_1$ & $[e_1,e_3]=e_1$, $[e_2,e_3]=a e_2$ \\
$4$ & $s_{3,3}(b)\oplus n_{1,1}$ & $A^b_{3.5}\oplus A_1$ & $[e_1,e_3]=b e_1-e_2$, $[e_2,e_3]=e_1+b e_2$ \\
$4$ & $\mathrm{sl}(2,\mathbb R)\oplus n_{1,1}$ & $\mathrm{sl}(2,\mathbb R)\oplus A_1$ & $[e_1,e_2]=e_1$, $[e_2,e_3]=e_3$, $[e_1,e_3]=2e_2$ \\
$4$ & $\mathrm{so}(3)\oplus n_{1,1}$ & $\mathrm{so}(3)\oplus A_1$ & $[e_2,e_3]=e_1$, $[e_3,e_1]=e_2$, $[e_1,e_2]=e_3$ \\
\bottomrule
\end{tabulary}
\caption{The decomposable real Lie algebras of dimension at most four.}
\label{decomposablelist}
\end{table}

The Abelian decomposable algebras are especially simple (table \ref{decomposableabelian}). Since every subalgebra of an Abelian Lie algebra is an ideal, the only faithful homogeneous Shirokov realization of $mn_{1,1}=mA_1$ is the generic one. Popovych's tables contain further faithful Abelian realizations, but these are intransitive and therefore do not arise from the homogeneous Shirokov construction.

\begin{table}[H]
\centering
\small
\begin{tabulary}{\linewidth}{LLCLCLR}
\toprule
Algebra & Subalgebra & dim. & Kernel & Faithful? & Shirokov realization & Popovych realization \\
\midrule
$2n_{1,1}$ & $\{0\}$ & $2$ & $\{0\}$ & yes & $e_1=\partial_1$, $e_2=\partial_2$ & $R(2A_1,1)$ \\
\addlinespace
$3n_{1,1}$ & $\{0\}$ & $3$ & $\{0\}$ & yes & $e_1=\partial_1$, $e_2=\partial_2$, $e_3=\partial_3$ & $R(3A_1,1)$ \\
\addlinespace
$4n_{1,1}$ & $\{0\}$ & $4$ & $\{0\}$ & yes & $e_1=\partial_1$, $e_2=\partial_2$, $e_3=\partial_3$, $e_4=\partial_4$ & $R(4A_1,1)$ \\
\bottomrule
\end{tabulary}
\caption{Faithful homogeneous Shirokov realizations of the decomposable Abelian algebras.}
\label{decomposableabelian}
\end{table}

For non-Abelian direct sums, the generic rows are obtained by adding independent variables to the generic rows of the summands. Table \ref{decomposablegeneric} records these generic rows. In the rows $A_{3.i}\oplus A_1$, the symbol $R_{\mathrm{gen}}(A_{3.i})$ means the generic realization of the corresponding three-dimensional algebra already displayed in the main text, with the additional central generator realized as $e_4=\partial_4$.

\begin{table}[H]
\centering
\small
\begin{tabulary}{\linewidth}{LLR}
\toprule
Algebra & Generic Shirokov realization & Popovych realization \\
\midrule
$s_{2,1}\oplus n_{1,1}$ & $e_1=\partial_1$, $e_2=x_1\partial_1+\partial_2$, $e_3=\partial_3$ & $R(A_{2.1}\oplus A_1,1)$ \\
\addlinespace
$s_{2,1}\oplus 2n_{1,1}$ & $e_1=\partial_1$, $e_2=x_1\partial_1+\partial_2$, $e_3=\partial_3$, $e_4=\partial_4$ & $R(A_{2.1}\oplus 2A_1,1)$ \\
\addlinespace
$2s_{2.1}$ & $e_1=\partial_1$, $e_2=x_1\partial_1+\partial_2$, $e_3=\partial_3$, $e_4=x_3\partial_3+\partial_4$ & $R(2A_{2.1},1)$ \\
\addlinespace
$n_{3,1}\oplus n_{1,1}$ & $R_{\mathrm{gen}}(n_{3,1})$, $e_4=\partial_4$ & $R(A_{3.1}\oplus A_1,1)$ \\
\multicolumn{3}{c}{(continued on next page)}\end{tabulary}\end{table}\begin{table}[H]\centering\small\begin{tabulary}{\linewidth}{LLR} Algebra & Generic Shirokov realization & Popovych realization \\ \midrule \multicolumn{3}{c}{(continued from previous page)} \\
$s_{3,2}\oplus n_{1,1}$ & $R_{\mathrm{gen}}(s_{3,2})$, $e_4=\partial_4$ & $R(A_{3.2}\oplus A_1,1)$ \\
\addlinespace
$s_{3,1}(1)\oplus n_{1,1}$ & $R_{\mathrm{gen}}(s_{3,1}(1))$, $e_4=\partial_4$\hphantom{xxxxxxxxxxxxxxxxxxxx} & $R(A_{3.3}\oplus A_1,1)$ \\
\addlinespace
$s_{3,1}(a)\oplus n_{1,1}$ & $R_{\mathrm{gen}}(s_{3,1}(a))$, $e_4=\partial_4$ & $R(A^a_{3.4}\oplus A_1,1)$ \\
\addlinespace
$s_{3,3}(b)\oplus n_{1,1}$ & $R_{\mathrm{gen}}(s_{3,3}(b))$, $e_4=\partial_4$ & $R(A^b_{3.5}\oplus A_1,1)$ \\
\addlinespace
$\mathrm{sl}(2,\mathbb R)\oplus n_{1,1}$ & $R_{\mathrm{gen}}(\mathrm{sl}(2,\mathbb R))$, $e_4=\partial_4$ & $R(\mathrm{sl}(2,\mathbb R)\oplus A_1,1)$ \\
\addlinespace
$\mathrm{so}(3)\oplus n_{1,1}$ & $R_{\mathrm{gen}}(\mathrm{so}(3))$, $e_4=\partial_4$ & $R(\mathrm{so}(3)\oplus A_1,1)$ \\
\bottomrule
\end{tabulary}
\caption{Generic faithful homogeneous Shirokov realizations of non-Abelian decomposable algebras.}
\label{decomposablegeneric}
\end{table}

We finish this section with the smallest non-Abelian decomposable example, $s_{2,1}\oplus n_{1,1}$. This example (see table \ref{a21plusa1}) shows the role of Goursat-type subalgebras. The Popovych realization $R(A_{2.1}\oplus A_1,2)$ is faithful but intransitive, hence it is not included in table \ref{a21plusa1}. The remaining faithful Popovych rows are homogeneous and appear from the subalgebras listed below.

\begin{table}[H]
\centering
\small
\begin{tabularx}{\linewidth}{lll}
\toprule
Algebra &  dim. & Nonzero commutation relations \\
$s_{2,1}\oplus n_{1,1}$ &   $3$ & $[e_1,e_2]=e_1$, $e_3$ central \\
\bottomrule
\end{tabularx}
\begin{tabularx}{\linewidth}{ll}
Patera-Winternitz & Popovych \\
$A_2\oplus A_1$ &   $A_{2.1}\oplus A_1$ \\
$[p_1,p_2]=p_2$ & \\
$p_1=-e_2$, $p_2=e_1$, $p_3=e_3$ &  \\
\bottomrule
\end{tabularx}
\medskip
\begin{tabulary}{\linewidth}{LLCLCLR}
\toprule
Subalgebra & Complement & dim. & Kernel & Faithful? & Shirokov realization & Popovych realization \\
\midrule
$\{0\}$
& $\langle e_1,e_2,e_3\rangle$
& $3$ & $\{0\}$ & yes
& $e_1=\partial_1$, $e_2=x_1\partial_1+\partial_2$, $e_3=\partial_3$
& $R(A_{2.1}\oplus A_1,1)$ \\
\addlinespace
$\langle e_2\rangle$
& $\langle e_1,e_3\rangle$
& $2$ & $\{0\}$ & yes
& $e_1=\partial_1$, $e_2=x_1\partial_1$, $e_3=\partial_2$
& $R(A_{2.1}\oplus A_1,3)$ \\
\addlinespace
$\langle e_1-e_3\rangle$
& $\langle e_1,e_2\rangle$
& $2$ & $\{0\}$ & yes
& $e_1=\partial_1$, $e_2=x_1\partial_1+x_2\partial_2$, $e_3=x_2\partial_1$
& $R(A_{2.1}\oplus A_1,4)$ \\
\bottomrule
\end{tabulary}
\caption{The faithful Shirokov realizations of $s_{2,1}\oplus n_{1,1}$.}
\label{a21plusa1}
\end{table}

For example, in the last row of table \ref{a21plusa1}, the isotropy subalgebra at the point $(0,1)$ is $\langle e_1-e_3\rangle$. This row is not a direct product row; it is a simple instance of the Goursat phenomenon for subalgebras of direct sums.

\end{appendices}

\section*{Acknowledgements}
The author is deeply grateful to Maryna Nesterenko for drawing his attention to Shirokov's method, which served as the primary motivation for this work.

\printbibliography

\end{document}